\documentclass[aps,prd,preprint,groupedaddress,showpacs]{revtex4}

\usepackage{epsfig}
\usepackage{graphics}
\usepackage{slashed}
\usepackage{color}
\begin{document}

\leftmargin -2cm
\def\choosen{\atopwithdelims..}

\boldmath
\title{Evidence in favor of Single Parton Scattering mechanism \\ in $\Upsilon$ and $D$ associated production at the LHC} \unboldmath

\author{\firstname{A.V.}\surname{Karpishkov}} \email{karpishkoff@gmail.com}
\author{\firstname{M.A.}
\surname{Nefedov}} \email{nefedovma@gmail.com}
\author{\firstname{V.A.}\surname{Saleev}} \email{saleev@samsu.ru}

\affiliation{Samara National Research University, Moskovskoe Shosse,
34, 443086, Samara, Russia}

\begin{abstract} Associated production of prompt $\Upsilon(1S)$ and $D^{+,0}$ mesons
has been proposed as a golden channel for the search of Double
Parton Scattering, because Single Parton Scattering contribution to
the cross-section is believed to be negligible on a basis of leading
order calculations in the Collinear Parton Model. We study this
process in the leading order approximation of the Parton
Reggeization Approach. Hadronization of $b\bar{b}$-pair into
bottomonium states is described within framework of the
NRQCD-factorization approach while production of $D$ mesons is
described in the fragmentation model with scale-dependent
fragmentation functions. We have found good agreement with LHCb data
for various normalized differential distributions, except for the
case of spectra on azimuthal angle differences at the small
$\Delta\varphi$ values. Crucially, the total cross-section in our
Single Parton Scattering model accounts for {more than} one half of
observed cross-section, thus dramatically shrinking the room for
Double Parton Scattering mechanism.
\end{abstract}



\maketitle

\section{Introduction}
\label{sec:Int} Double Parton Scattering (DPS) mechanism attracts a
lot of attention, both theoretically and experimentally. Recent
advances in theory of DPS include computation~\cite{Diehl:2019rdh}
of NLO corrections to $1\to 2$ splitting functions, contributing to
the scale-evolution of two-parton collinear Parton Diastrbution
Functions (PDFs) and filling remaining gaps in the proof of
TMD-factorization theorem for double Drell-Yan
process~\cite{Diehl:2018wfy}.

  While for single-scale observables,
such as total inclusive cross-section of some hard process with hard
scale $Q$, the DPS contribution is always suppressed by powers of
$\Lambda^2/Q^2$, where $\Lambda$ is some typical hadronic scale, the
DPS contribution can be comparable with Single Parton Scattering
contribution for the differential distributions in some regions of
phase space~\cite{Diehl:2011yj}. In the DPS picture, longitudinal
and transverse-momentum correlations of small-$x$ partons,
participating in two independent hard scatterings, are quickly
washed-out by effects of scale evolution of two-parton
PDFs~\cite{Elias:2017flu}, so that double parton PDF effectively
factorizes into a product of usual PDFs, leading to the
``pocket-formula'' description of DPS as:
\begin{equation}
\sigma^{}_{\rm DPS}\simeq \frac{\sigma^{(1)}_{\rm SPS}\times
\sigma^{(2)}_{\rm SPS}}{\sigma_{\rm eff}}, \label{Eq:pocket-formula}
\end{equation}
where $\sigma^{}_{\rm DPS}$ is the DPS contribution to the
production cross-section of the pair of objects,
$\sigma^{(1,2)}_{\rm SPS}$ is the Single Parton Scattering (SPS)
cross-section of the production of one final-state object and
$\sigma_{\rm eff}$ is the non-perturbative ``effective
cross-section'' which is typically assumed to be universal,
energy-independent parameter. Eq.~(\ref{Eq:pocket-formula}) leads to
flat distributions for observables correlating momenta of particles
produced in independent hard scatterings, in particular, for the
azimuthal angle $\Delta\varphi$ or rapidity difference $\Delta y$
between momenta of components of the pair. In contrast, the SPS
contribution to differential distributions in the processes of
pair-production of jets, vector bosons or mesons containing heavy
flavors, typically decreases rather steeply with increasing $|\Delta
y|$ or $\Delta\varphi$ varying from $\Delta\varphi=\pi$ down to
zero. Therefore, an overshoot of experimentally-observed
cross-section over best available SPS predictions in the region
$\Delta\varphi\to 0$ or $|\Delta y|\gg 1$ can be interpreted as a
signature of DPS.


Another way to probe the DPS mechanism is to use the fact that due
to Eq.~(\ref{Eq:pocket-formula}) the DPS cross-section grows with
energy roughly as a square of SPS cross-section, so at sufficiently
high energy the power-supression by hard scale will be compensated
by quickly growing cross-section and the DPS contribution will
dominate. To observe this effect at available energies, one chooses
the process, {for which} SPS cross-section is as small as possible,
due to suppression by high power of $\alpha_s$. The hard scale also
should be small in comparison to the collision energy, so that one
stays in the low-$x$ region, where Eq.~(\ref{Eq:pocket-formula}) is
applicable. Pair production of mesons containing $c$ and $b$ quarks
seems to be an ideal playground for this kind of studies. But one
have to anticipate, that due to low hard scales and involvement of
relatively poorly understood physics of hadronization, the SPS
cross-section could receive unexpectedly large higher-order QCD
corrections, which could invalidate the assumption of negligibility
of the SPS contribution. Moreover, at low-$x$ plenty of phase space
is available for emission of additional relatively hard partons,
which will broaden the distributions in variables like
$\Delta\varphi$ and $\Delta y$, thus mimicking the effects of DPS
and making the extraction of the DPS signal from differential
distributions more model-dependent.

In Ref.~\cite{LHCb_ups_D} the strategy described in previous
paragraph has been applied by LHCb-Collaboration for the search of
DPS contribution in the process of associated production of prompt
bottomonim $\Upsilon(1S)$ and open-charm mesons $D^0$ or $D^+$ in
forward-rapidity region in $pp$-collisions with center-of-mass
energies $\sqrt{S}=7$ and $8$ TeV at the Large Hadron Collider. The
leading order (LO, $O(\alpha_s^4)$ in this case) calculations in
Collinear Parton Model~\cite{Likhoded2015,Likhoded2016} and
Color-Singlet approximation for the hadronization of $b\bar{b}\to
\Upsilon$ lead to a very small SPS total cross-section of this
process, compared with the cross-section of inclusive production of
$\Upsilon$-mesons, calculated in a same approximation: $R_{\rm
SPS}^{\rm (LO)}=\sigma^{(LO,\ CS)}_{SPS}(\Upsilon + c
\bar{c})/\sigma^{(LO,\ CS)}_{SPS}(\Upsilon)=(0.2-0.6)\%$, while
experimentally this ratio reaches almost $8\%$ at $\sqrt{S}=8$
TeV~\cite{LHCb_ups_D}. Such a large discrepancy was interpreted in
Ref.~\cite{LHCb_ups_D} as a clear signal of DPS mechanism. This
interpretation was supported by the fact, that shapes of
differential distributions measured in Ref.~\cite{LHCb_ups_D} can be
reasonably described by simple Monte-Carlo calculations based on
measured inclusive cross-sections of production of $\Upsilon$ and
$D$-mesons and ``pocket-formula''~(\ref{Eq:pocket-formula}).

In the present paper we reconsider the estimation of SPS
contribution to above-described process, adding a few effects which
{were missed} in calculations of
Refs.~\cite{Likhoded2015,Likhoded2016}. First, we approximately take
into account higher-order corrections coming from Initial-State
Radiation (ISR) effects, by virtue of Parton Reggeization Approach
(PRA)~\cite{NSS2013,NKS2017}, which allows to factorize most
significant part of ISR corrections into unintegrated Parton
Distribution Functions (unPDFs) consistently with QCD
gauge-invariance. The latter fact allows us to take into account the
contributions to the $b\bar{b}\to \Upsilon$ coming from color-octet
$b\bar{b}$-pair in a framework of nonrelativistic QCD (NRQCD)
factorization approach~\cite{NRQCD}. An importance of color-octet
contributions for bottomonium production in PRA has been
demonstrated by us in Ref.~\cite{NSSUpsilon}. And finally, we
consistently take into account parton to $D$-meson hadronization,
using the scale-dependent fragmentation functions, obtained in a
global analysis of open-charm hadron production in
$e^+e^-$-annihilation in Ref.~\cite{KKSS}. Counter-intuitively, not
only $c\to D$ but also $g\to D$ fragmentation plays an important
role in the associated $\Upsilon$ and $D$ production. Gluon
fragmentation has also been found to be important for the same-sign
$DD$-pair production in Refs.~\cite{PLB2016,PRD2016DD}. In total and
taking into account the uncertainties, our hybrid
PRA+NRQCD+fragmentation model can account {more than} a half of
experimentally observed cross-section for $\Upsilon D$ pair
production. The shapes of all differential distributions in our SPS
model also turn out to reproduce experimental data rather well,
except for the shape of $\Delta\varphi$-spectrum which has a
puzzling ``upside-down'' behavior, un-explainable even by the DPS
model (\ref{Eq:pocket-formula}). Therefore, we show that radiative
corrections to the SPS cross-section of the process under
consideration are large and mimicking the behavior of DPS, thus
weakening the case for DPS-dominance in this process.

The paper is organized as follows. We describe the basics of PRA in
the Sec.~\ref{sec:PRA}. In the Sec.~\ref{sec:model} we present our
model  of $\Upsilon(1S)D^{0,+}$ pair production. Then we concentrate
on the numerical results and comparison with experimental data of
the Ref.~\cite{LHCb_ups_D} in the Sec.~\ref{sec:results}. Finally,
we summarize our conclusions in the Sec.~\ref{sec:conclusions}.

\section{Parton Reggeization Approach}
\label{sec:PRA} The brief description of LO approximation of PRA is
presented below. More details can be found in Ref.~\cite{NKS2017},
the development of PRA in the next-to-leading order (NLO)
approximation is further discussed in \cite{PRANLO1,PRANLO2}. The
main ingredients of PRA are $k_T-$dependent factorization formula,
unintegrated parton distribution functions (unPDF's) and
gauge-invariant amplitudes with off-shell initial-state partons,
derived using the Lipatov's Effective Field Theory (EFT) of
Reggeized gluons~\cite{Lipatov95} and Reggeized
quarks~\cite{LipatovVyazovsky}.

Factorization formula of PRA in LO approximation for the process
$p+p\to {\cal Y}+ X$, can be obtained from factorization formula of
the CPM for the auxiliary hard subprocess $g+g\to g + {\cal Y}+ g$.
In the Ref.~\cite{NKS2017} the modified Multi-Regge Kinematics
(mMRK) approximation for the auxiliary amplitude is constructed,
which correctly reproduces the Multi-Regge and collinear limits of
corresponding QCD amplitude. This mMRK-amplitude has $t$-channel
factorized form, which allows one to rewrite the cross-section of
auxiliary subprocess in a $k_T$-factorized form:
  \begin{eqnarray}
  d\sigma &=& \int\limits_0^1 \frac{dx_1}{x_1} \int \frac{d^2{\bf q}_{T1}}{\pi} \tilde{\Phi}_g(x_1,t_1,\mu^2)
\int\limits_0^1 \frac{dx_2}{x_2} \int \frac{d^2{\bf q}_{T2}}{\pi}
\tilde{\Phi}_g(x_2,t_2,\mu^2)\cdot d\hat{\sigma}_{\rm PRA},
\label{eqI:kT_fact}
  \end{eqnarray}
where $t_{1,2}=-{\bf q}_{T1,2}^2$, the off-shell partonic
cross-section $\hat\sigma_{\rm PRA}$ in PRA is determined by squared
PRA amplitude, $\overline{|{\cal A}_{PRA}|^2}$. Despite the fact
that four-momenta ($q_{1,2}$) of partons in the initial state of
amplitude  ${\cal A}_{PRA}$ are off-shell ($q_{1,2}^2=-t_{1,2}<0$),
the PRA hard-scattering amplitude is {\it gauge-invariant} because
the initial-state off-shell gluons are treated as Reggeized gluons
($R$) in a sense of gauge-invariant EFT for QCD processes in
Multi-Regge Kinematics(MRK), introduced by L.N. Lipatov
in~\cite{Lipatov95}. The Feynman rules of this EFT are written down
in Ref.~\cite{AntonovLipatov}.

 The tree-level "unintegrated PDFs"~ (unPDFs) $\tilde{\Phi}_g(x_{1,2},t_{1,2},\mu^2)$ in Eq. (\ref{eqI:kT_fact})
 are equal to the convolution
of the collinear PDF $f_g(x,\mu^2)$ and
Dokshitzer-Gribov-Lipatov-Altarelli-Parisi (DGLAP) splitting
function $P_{gg}(z)$ with the factor $1/t_{1,2}$. Consequently, the
cross-section (\ref{eqI:kT_fact}) with such unPDFs contains the
collinear divergence at $t_{1,2}\to 0$ and infrared (IR) divergence
at $z_{1,2}\to 1$. To regularize the latter, we observe, that the
mMRK expression gives a reasonable approximation for the exact
matrix element only in the rapidity-ordered part of the phase-space
$y_{g_1}>y_{\cal Y}>y_{g_2}$. From this requirement, the following
cutoff on $z_{1,2}$ can be derived: $z_{1,2}<
1-\Delta_{KMR}(t_{1,2},\mu^2),$
  where $\Delta_{KMR}(t,\mu^2)=\sqrt{t}/(\sqrt{\mu^2}+\sqrt{t})$ is the Kimber-Martin-Ryskin (KMR) cutoff function~\cite{KMR}, and we have
 taken into account that $\mu^2\sim M_{T{\cal Y}}^2$. The collinear
singularity at $t_{1,2}\to 0$ is regularized by the Sudakov formfactor:
  \begin{equation}
  T_i(t,\mu^2)=\exp\left[ - \int\limits_t^{\mu^2} \frac{dt'}{t'} \frac{\alpha_s(t')}{2\pi}
\sum\limits_{j=q,\bar{q},g} \int\limits_0^{1} dz\ z\cdot P_{ji}(z)
\theta\left(1-\Delta_{KMR}(t',\mu^2) - z\right)  \right],
\label{eq:Sudakov}
  \end{equation}
  which resums doubly-logarithmic corrections $\sim\log^2 (t/\mu^2)$ in the Leading-Logarithmic-Approximation.

The final form of our unPDF in PRA is:
  \begin{equation}
  \Phi_i(x,t,\mu^2) = \frac{T_i(t,\mu^2)}{t} \frac{\alpha_s(t)}{2\pi} \sum_{j=q,\bar{q},g}
   \int\limits_x^{1} dz\ P_{ij}(z)\cdot \frac{x}{z}f_{j}\left(\frac{x}{z},t \right)\cdot \theta
   \left(1-\Delta_{KMR}(t,\mu^2)-z \right), \label{eqI:KMR}
  \end{equation}
which coincides with Kimber, Martin and Ryskin unPDF~\cite{KMR}. The
KMR unPDF is actively used in the phenomenological studies employing
$k_T$-factorization, but to our knowledge, the derivation, presented
in~\cite{NKS2017} is the first systematic attempt to clarify it's
relationships with MRK limit of the QCD amplitudes.

In contrast to most of studies in the $k_T$-factorization, the
gauge-invariant matrix elements with off-shell initial-state partons
(Reggeized quarks and gluons) from Lipatov's EFT~\cite{Lipatov95,
LipatovVyazovsky} allow one to study arbitrary processes involving
non-Abelian structure of QCD without violation of Slavnov-Taylor
identities due to the nonzero virtuality of initial-state partons.
This approach, together with KMR unPDFs gives stable and consistent
results in a wide range of phenomenological applications, which
include the description of the angular correlations of
dijets~\cite{NSS2013}, $b$-jets~\cite{SSbb}, charmed~\cite{PLB2016}
and bottom-flavored~\cite{NKS2017} mesons, as well as some other
examples.

A few years ago, the new approach to derive gauge-invariant
scattering amplitudes with off-shell initial-state partons for
high-energy scattering, using the spinor-helicity techniques and
BCFW-like recursion relations for such amplitudes has been
introduced in the Refs.~\cite{hameren1,hameren2, katie}. This
formalism for numerical generation of off-shell amplitudes is
equivalent to the Lipatov's EFT at the tree level, but for some
observables, e. g. related with production of heavy quarkonia, or
for the generalization of the formalism to NLO~\cite{PRANLO1, PRANLO2}, the use
of explicit Feynman rules and the structure of EFT is more
convenient.


\section{Model for $\Upsilon D$ pair production}
\label{sec:model}

In the Leading Order (LO), ($O(\alpha_s^3)$), in PRA plus
fragmentation model~\cite{Frag}, only gluon fragmentation
contributes to the process of associated production of bottomonium
and $D$-meson:
\begin{eqnarray}
R + R &\to& \Upsilon(3S,2S,1S)+ g(\to D), \label{RRSwave}\\
R + R &\to& \chi_b(2P,1P)+ g(\to D).\label{RRPwave}
\end{eqnarray}

One might worry, that the gain of one power of $\alpha_s$ will be
compensated by small numerical value of $g\to D$ fragmentation
function in comparison with $c\to D$ fragmentation and thus the
numerically leading contribution comes from the
Next-to-Leading-Order (NLO) ($O(\alpha_s^4)$) subprocesses:
\begin{eqnarray}
R + R &\to & \Upsilon(3S, 2S, 1S)+c(\to D)+\bar c, \label{2-3Swave} \\
R + R &\to & \chi_b(2P,1P) +c(\to D)+\bar c. \label{2-3Pwave}
\end{eqnarray}

We address this question in Sec.~\ref{sec:results} and show, that
contribution of $2\to 3$ subprocesses is actually subleading, so in
the final predictions for the cross-sections and spectra we take
into account only subprocesses (\ref{RRSwave}, \ref{RRPwave}).

\begin{figure}[]
\includegraphics[width=0.8\textwidth]{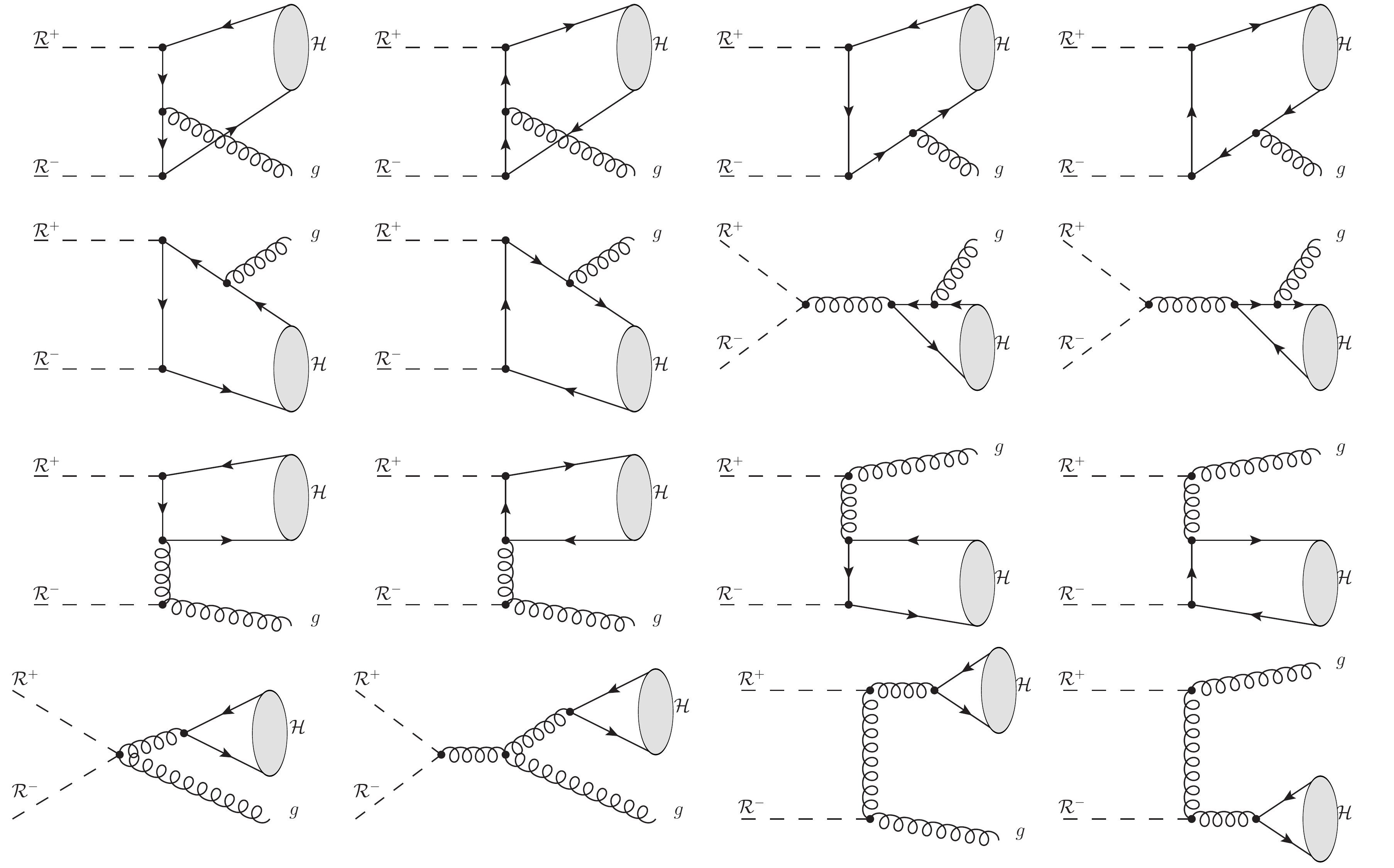}
 \caption{Feynman diagrams used for direct production of $S$ and $P-$wave
 bottomonia via color-singlet and color-octet intermediate states in processes $R+R\to b\bar b[^3S_1^{(1,8)}]+g$
 and $R+R\to b\bar b[^3P_{0,1,2}^{(1)}]+g$.}

 \label{fig-RRbbg}
\end{figure}

According to NRQCD factorization formalism~\cite{NRQCD}, final heavy
quarkonium can be produced via color-singlet and color-octet
intermediate states of $b\bar{b}$-pair. We use the set of
color-singlet and color-octet nonperturbative (long-distance) matrix
elements (NMEs or LDMEs), which has been obtained in the LO PRA plus
NRQCD-factorization approximation in the Ref.~\cite{NSSUpsilon} from
the fit of inclusive $p_T-$spectra of prompt $\Upsilon(nS)$-mesons,
measured by ATLAS, CMS and LHCb Collaborations at the LHC. For
reader's convenience, we collect these  NMEs in the
Table~\ref{TableI}. The color-octet contributions for the production
of $P$-wave bottomonia tend to be negligible in comparison with
color-singlet contributions~\cite{NSSUpsilon}, so in the $P$-wave
case we take into accunt only the singlet channel. Thus at the quark
level we are left with the following list of LO partonic
subprocesses:
\begin{eqnarray}
R + R &\to& b\bar b[^3S_1^{(1)}]+ g, \label{Ssinglet}\\
R + R &\to& b\bar b[^3S_1^{(8)}]+ g, \label{Soctet}\\
R + R &\to& b\bar b[^3P_{0,1,2}^{(1)}]+ g. \label{Psinglet}
\end{eqnarray}
The sets of Feynman diagrams of Lipatov's EFT which describe
subprocesses (\ref{Ssinglet}-\ref{Psinglet}) are shown in Fig.
\ref{fig-RRbbg}.
Squared off-shell amplitude of subprocess
(\ref{Ssinglet}) is well known, it was calculated many years ago in
the Ref.~\cite{KVS2006}. Squared off-shell amplitudes of subprocess
(\ref{Soctet}) and (\ref{Psinglet}) are calculated here for the
first time. However, corresponding on-shell squared amplitudes are
known \cite{ChoLeibovich} and they have been used for the test of
collinear limit of obtained PRA amplitudes with off-shell
initial-state partons.  To automatize the analytic calculations we
have implemented the Feynman rules of EFT~\cite{Lipatov95,
LipatovVyazovsky} as a model for FeynArts package~\cite{FeynArts}.
The further calculation of helicity amplitudes for $2\to 2$ and
$2\to 3$ processes has been performed using FeynCalc~\cite{FeynCalc}
package and for the numerical evaluation, amplitudes had been
squared, summed over colors and helicities and implemented as a
{\cal FORTRAN} code. Unfortunately, due to complicated dependence on
light-cone and transverse components of four-momenta of initial and
final-state particles, PRA amplitudes even for $2\to 2$ processes
tend to become prohibitively large for the journal publication. In
fact, in the case of $^3S_1^{(8)}$-subprocess (\ref{Soctet}) the
arithmetic with quadruple precision inside the squared-amplitude
routine is required to reach numerical stability of the calculation.

Considering the associated production of $\Upsilon(1S) D^{0,+}$
pairs, one should take into account both direct and feed-down
production of $\Upsilon(1S)$ via decays of higher-lying
$\Upsilon(3S,2S)$ and $\chi_b(2P,1P)$ states, which in turn are
produced directly or via decays of even excited states. Taking known
branching fractions of different decays from~\cite{PDG} and
considering cascades of up to three consequent decays we have
obtained the following cascade branching fractions for
$\Upsilon(1S)$-state: $\mbox{Br}(\Upsilon(3S)\to
\Upsilon(1S))=0.138$, $\mbox{Br}(\chi_{b2}(2P)\to
\Upsilon(1S))=0.114$, $\mbox{Br}(\chi_{b1}(2P)\to
\Upsilon(1S))=0.171$, $\mbox{Br}(\chi_{b0}(2P)\to
\Upsilon(1S))=0.023$, $\mbox{Br}(\Upsilon(2S)\to
\Upsilon(1S))=0.302$, $\mbox{Br}(\chi_{b2}(1P)\to
\Upsilon(1S))=0.191$, $\mbox{Br}(\chi_{b1}(1P)\to
\Upsilon(1S))=0.339$, $\mbox{Br}(\chi_{b0}(2P)\to
\Upsilon(1S))=0.018$, which are used in the calculation.

To describe the $D$-meson production we use the fragmentation model
in which the transition of gluon to the $D$ meson is described by
corresponding scale-dependent fragmentation function (FF)
$D_{g}(z,\mu^2)$ \cite{Frag}. Recently, the non-trivial role of
gluon fragmentation in associated production of same-sign $D-$meson
pairs at the LHC has been demonstrated by some of us~\cite{PLB2016}.
The latter process has been considered by some authors as clean
signal of DPS production mechanism \cite{DD_DPS}. In
Ref.~\cite{PLB2016} it has been shown that main mechanism of
same-sign $DD-$pair production is gluon to $D$ meson fragmentation
via production of gluon pair in the LO PRA subprocess $RR\to gg$.
The description of LHCb data \cite{LHCb_DD} has been archived
without hypothesis on large DPS contribution. In the present paper,
as well as in the Ref.~\cite{PLB2016} use universal scale-dependent
LO FFs of the Ref.~\cite{KKSS}, fitted to $e^+e^-$-annihilation data
from CERN LEP1.

\section{Results and discussion}
\label{sec:results} Here we discuss our numerical results obtained
for prompt $\Upsilon(1S)D^{0,+}$ pair production in $pp$-collisions
at energies $\sqrt{S}=7$ TeV and $\sqrt{S}=8$ TeV. We use the unPDFs
obtained by the KMR formula (\ref{eq:Sudakov}, \ref{eqI:KMR}) from
the LO collinear PDFs MSTW-2008~\cite{Martin:2009iq} and the
corresponding value of $\alpha_s(M_Z)=0.13939$. We set the
renormalization and factorization scales to
$\mu_R=\mu_F=\frac{\xi}{2}\left(\sqrt{M_{\Upsilon}^2+p_{T\Upsilon}^2}+\sqrt{M_{D}^2+p_{TD}^2}\right)$
where $\xi = 1$ for the central lines of our predictions, and we
vary $1/2 < \xi < 2$ to estimate the scale-uncertainty of our
prediction, which is shown in the figures by the gray band for the
curve corresponding to the sum of all contributions.

\begin{figure}[]
\includegraphics[width=0.5\textwidth]{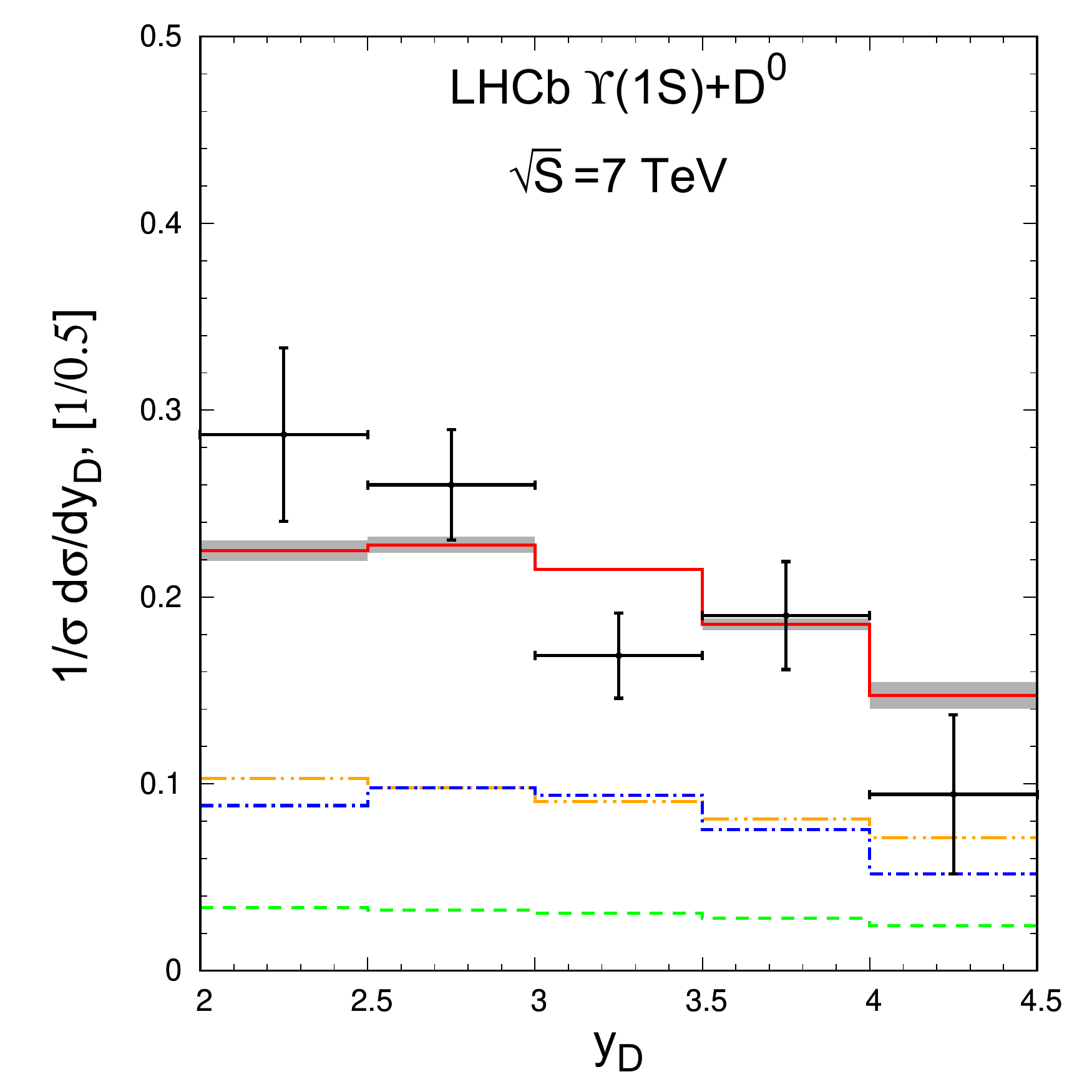}\includegraphics[width=0.5\textwidth]{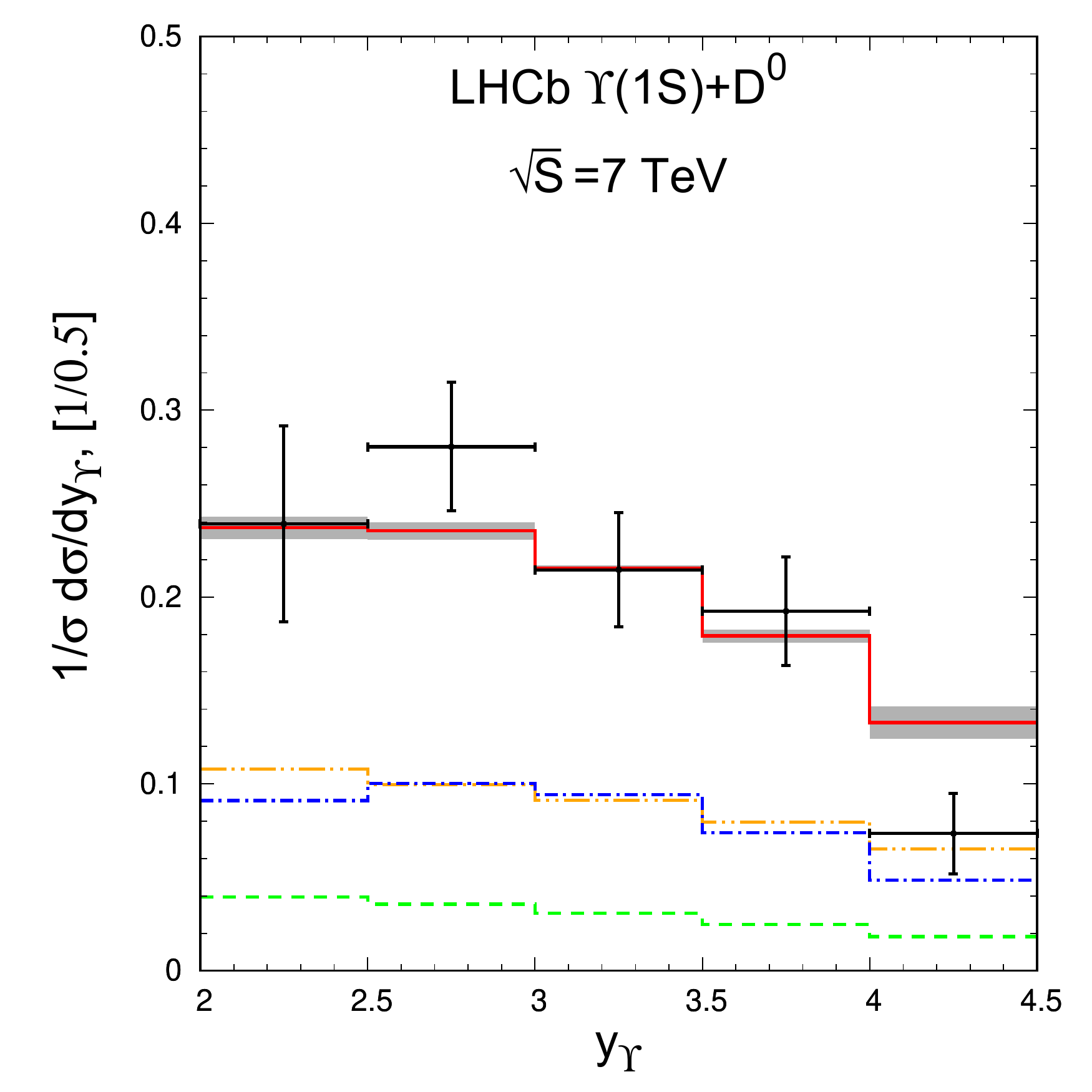}
\includegraphics[width=0.5\textwidth]{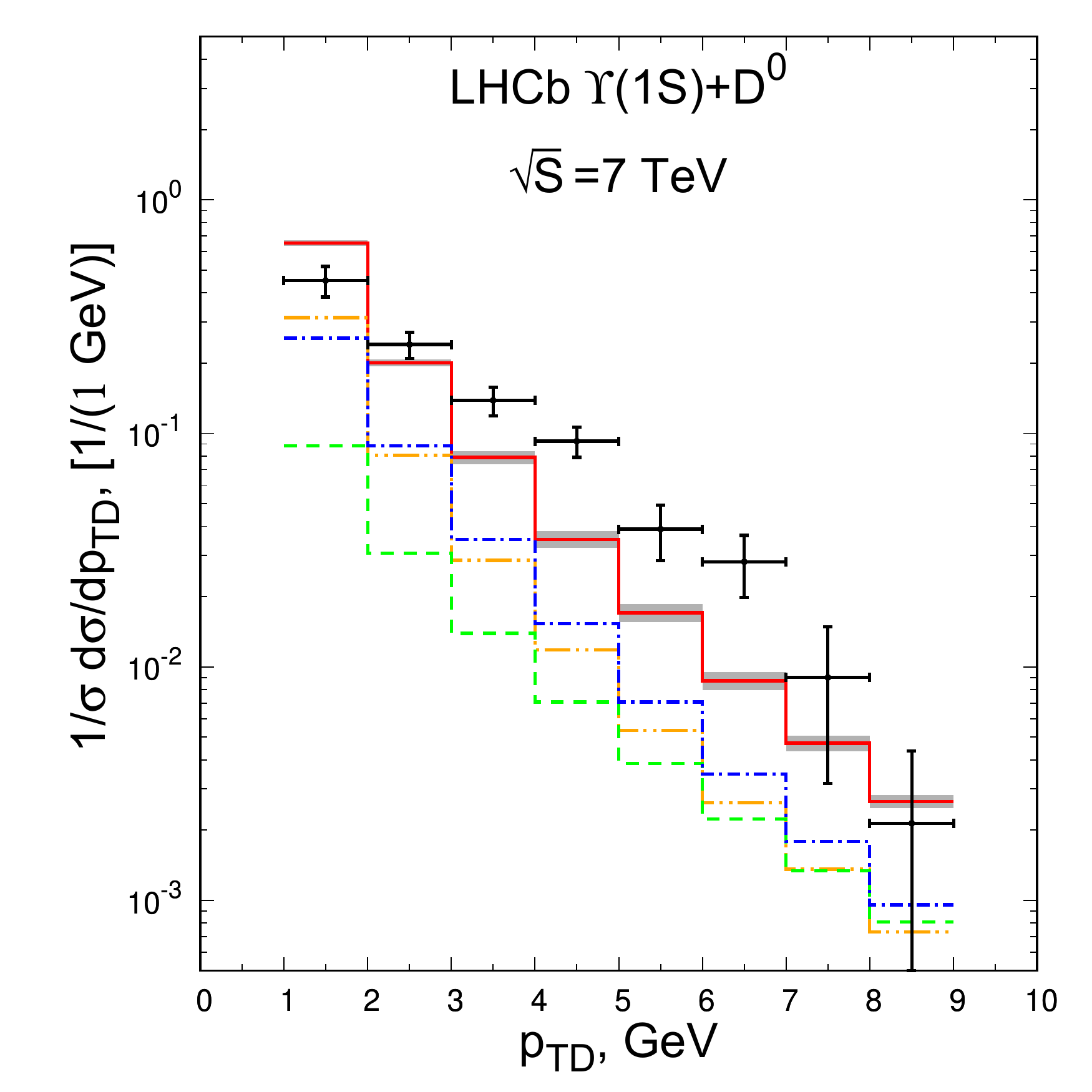}\includegraphics[width=0.5\textwidth]{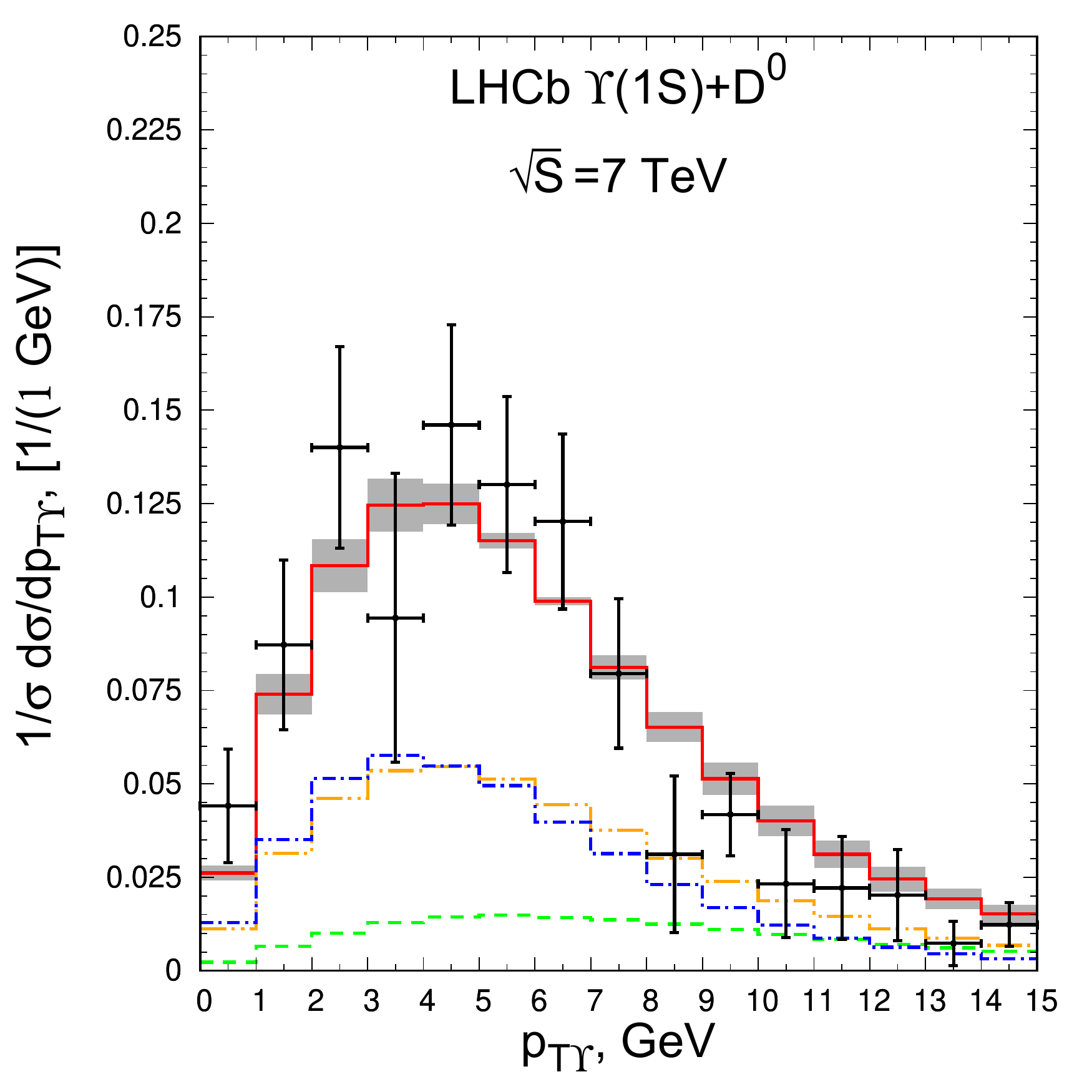}
 \caption{Transverse momentum and rapidity spectra of  $\Upsilon(1S)$ and $D^0$.
 The LHCb data~\cite{LHCb_ups_D} are taken at  $\sqrt{S}=7$ TeV, $2.0<y_{\Upsilon(D)}<4.5$, $0<p_{T\Upsilon}< 15$ GeV, and $1<p_{TD}< 20$ GeV.  Blue histograms are color-singlet
 contributions, green -- color-octet contributions, orange -- sum of feed-down contributions, red -- sum of all contributions.}
 \label{fig-1}
\end{figure}

\begin{figure}[]
\includegraphics[width=0.5\textwidth ]{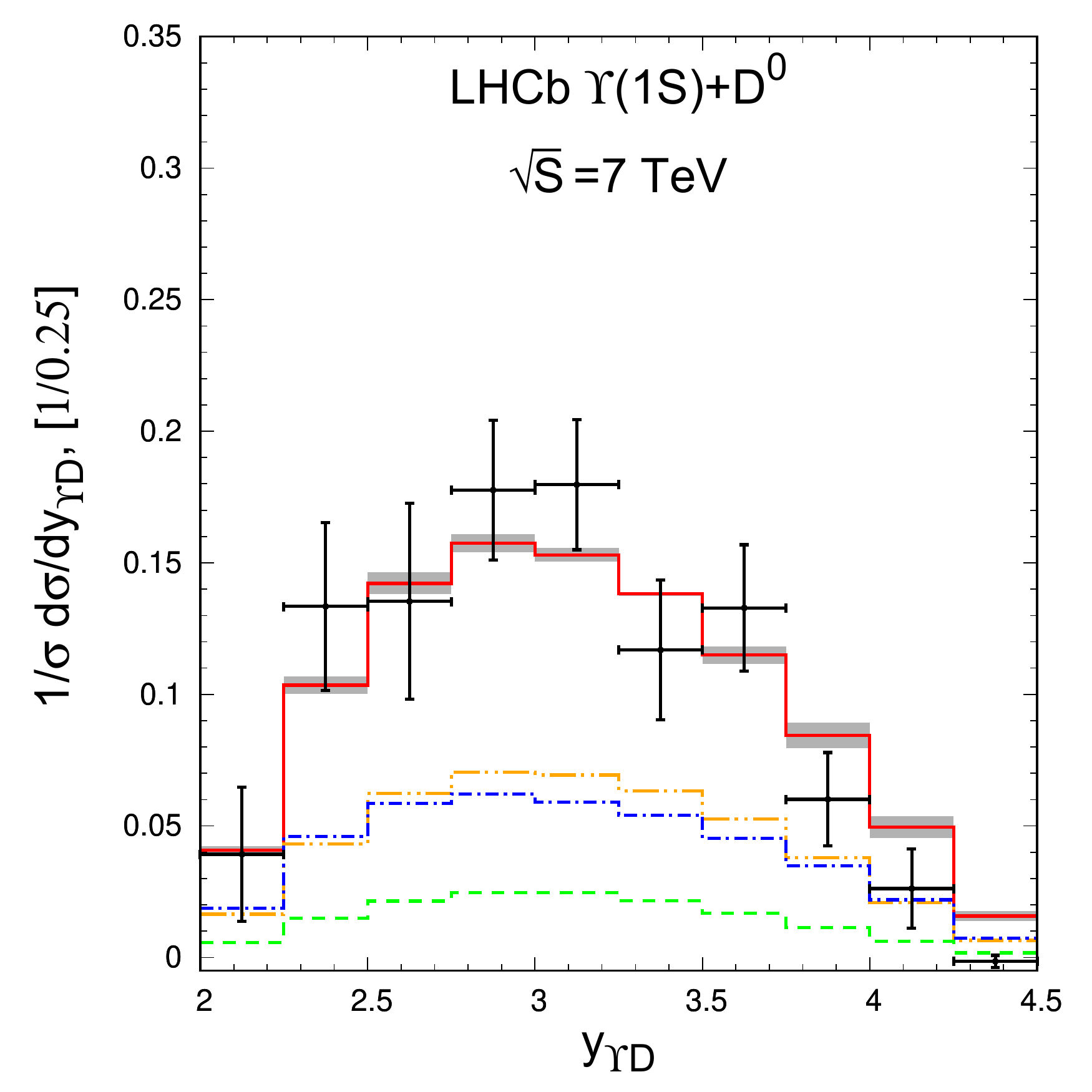}\includegraphics[width=0.5\textwidth  ]{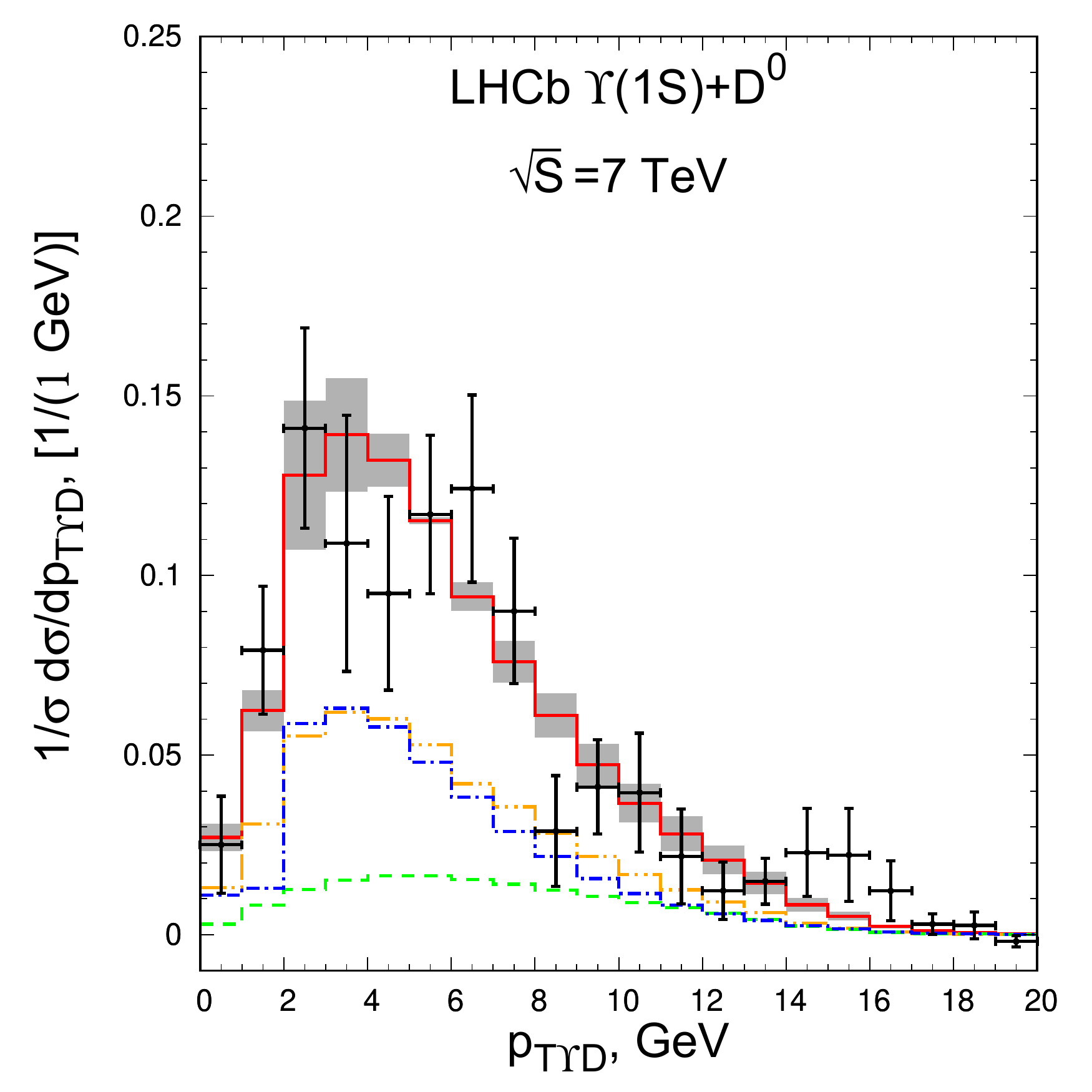}
\includegraphics[width=0.5\textwidth  ]{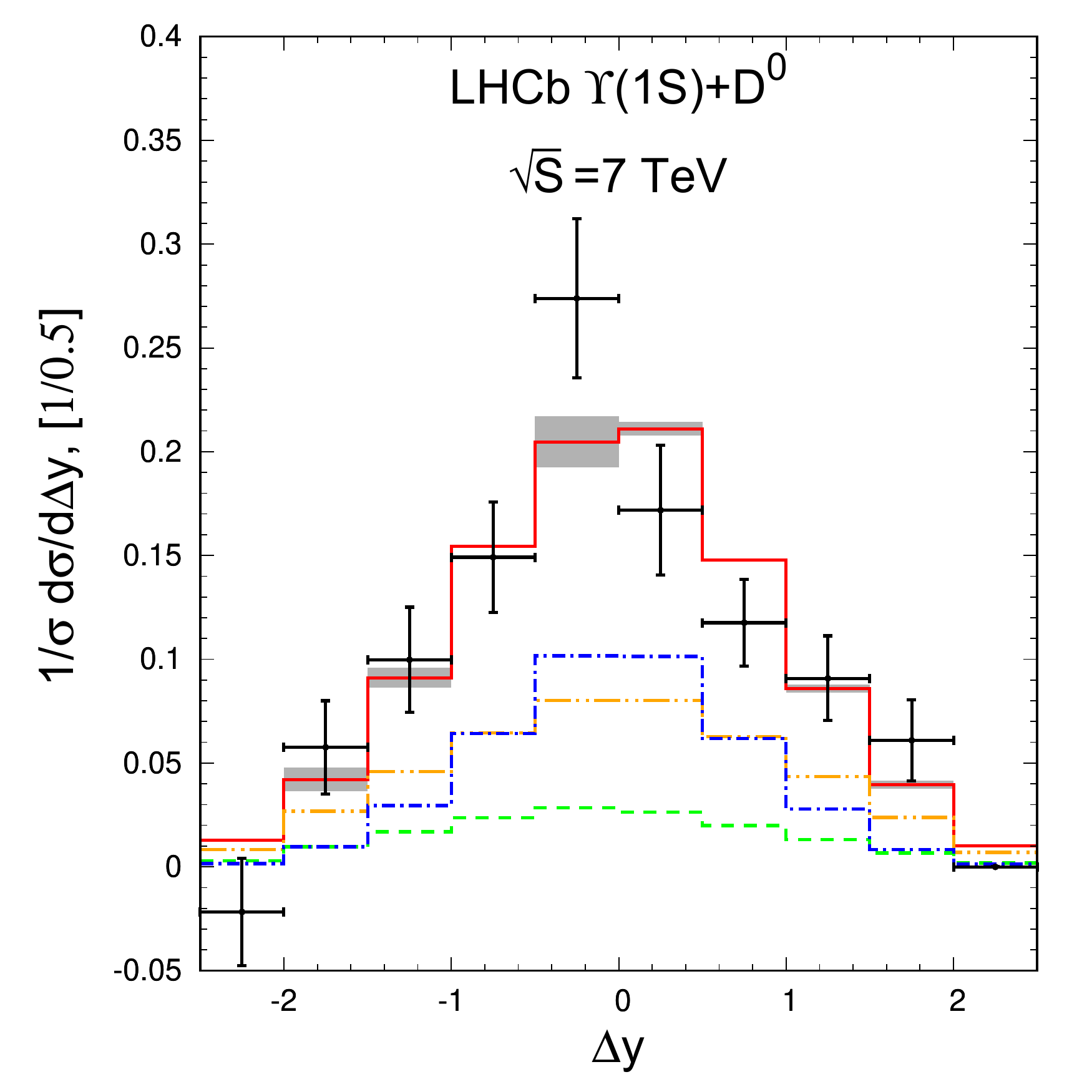}\includegraphics[width=0.5\textwidth  ]{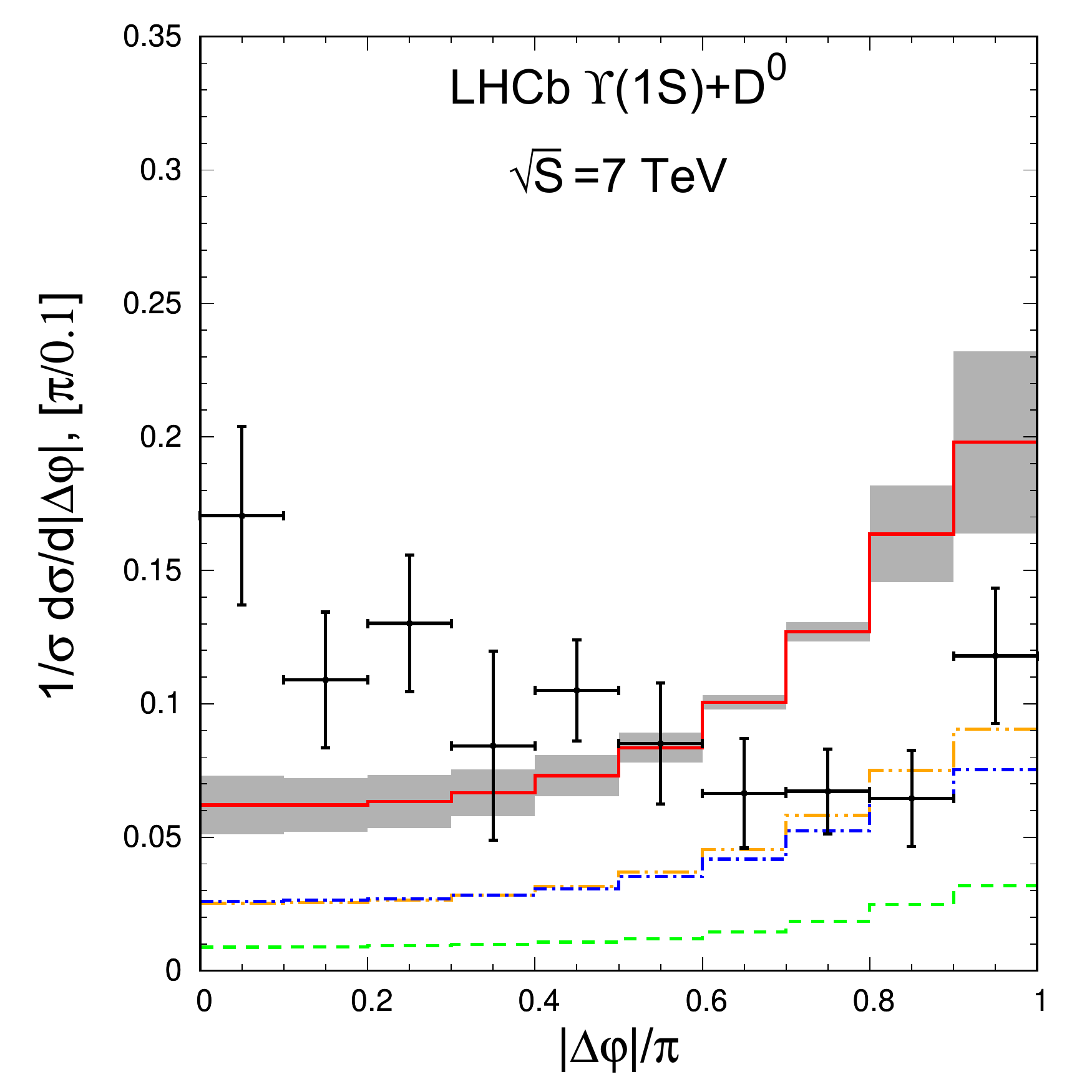}
 \caption{Top panel: transverse momentum and rapidity spectra of  $\Upsilon(1S)D^{0}$ pairs.
 Bottom panel: rapidity difference and azimuthal angle difference  spectra.  Histograms are the same as in the Fig. \ref{fig-1}. The LHCb data~\cite{LHCb_ups_D} are taken at
 $\sqrt{S}=7$ TeV, $2.0<y_{\Upsilon(D)}<4.5$, $0<p_{T\Upsilon}< 15$ GeV, and $1<p_{TD}< 20$ GeV.}
 \label{fig-2}
\end{figure}

\begin{figure}[]
\includegraphics[width=0.5\textwidth]{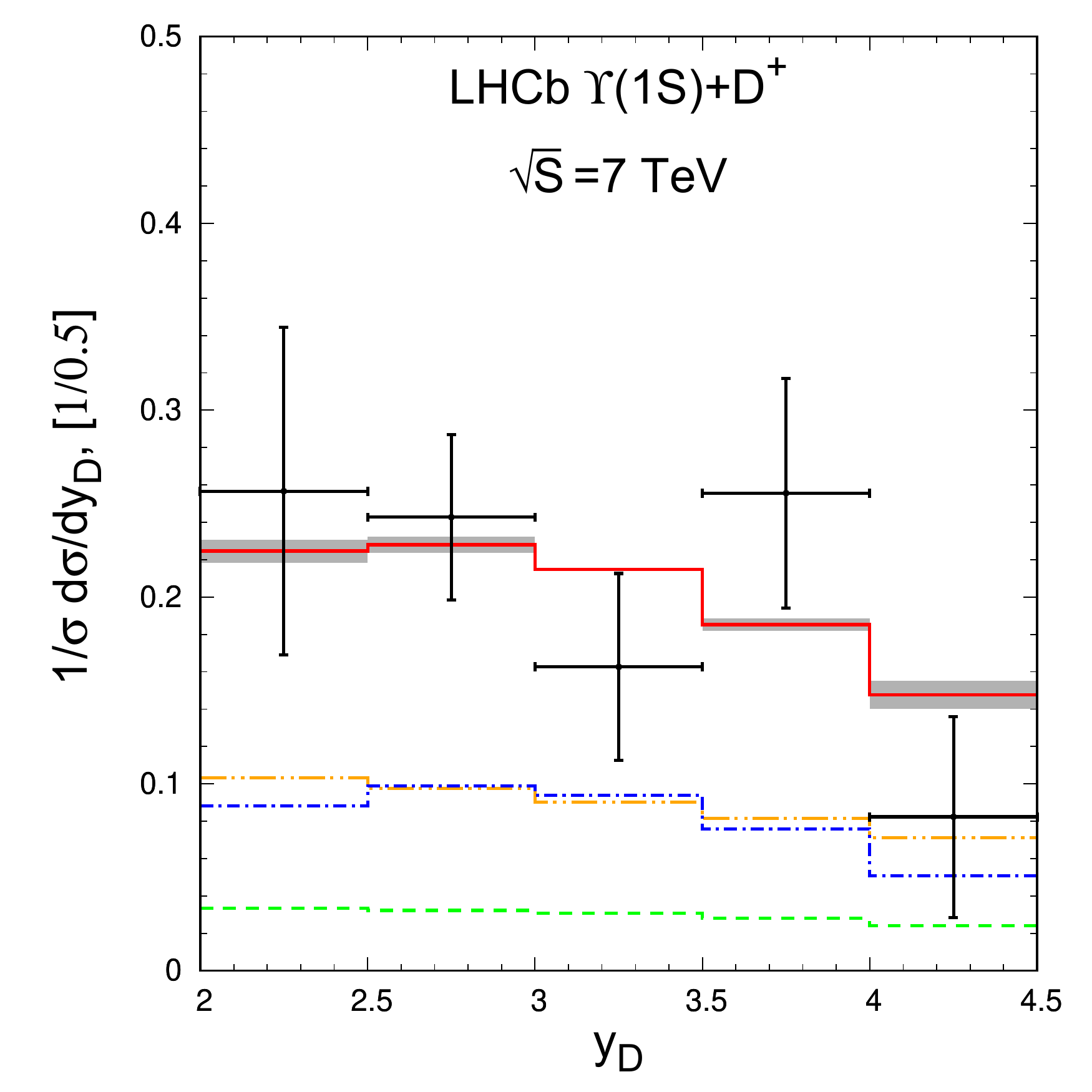}\includegraphics[width=0.5\textwidth ]{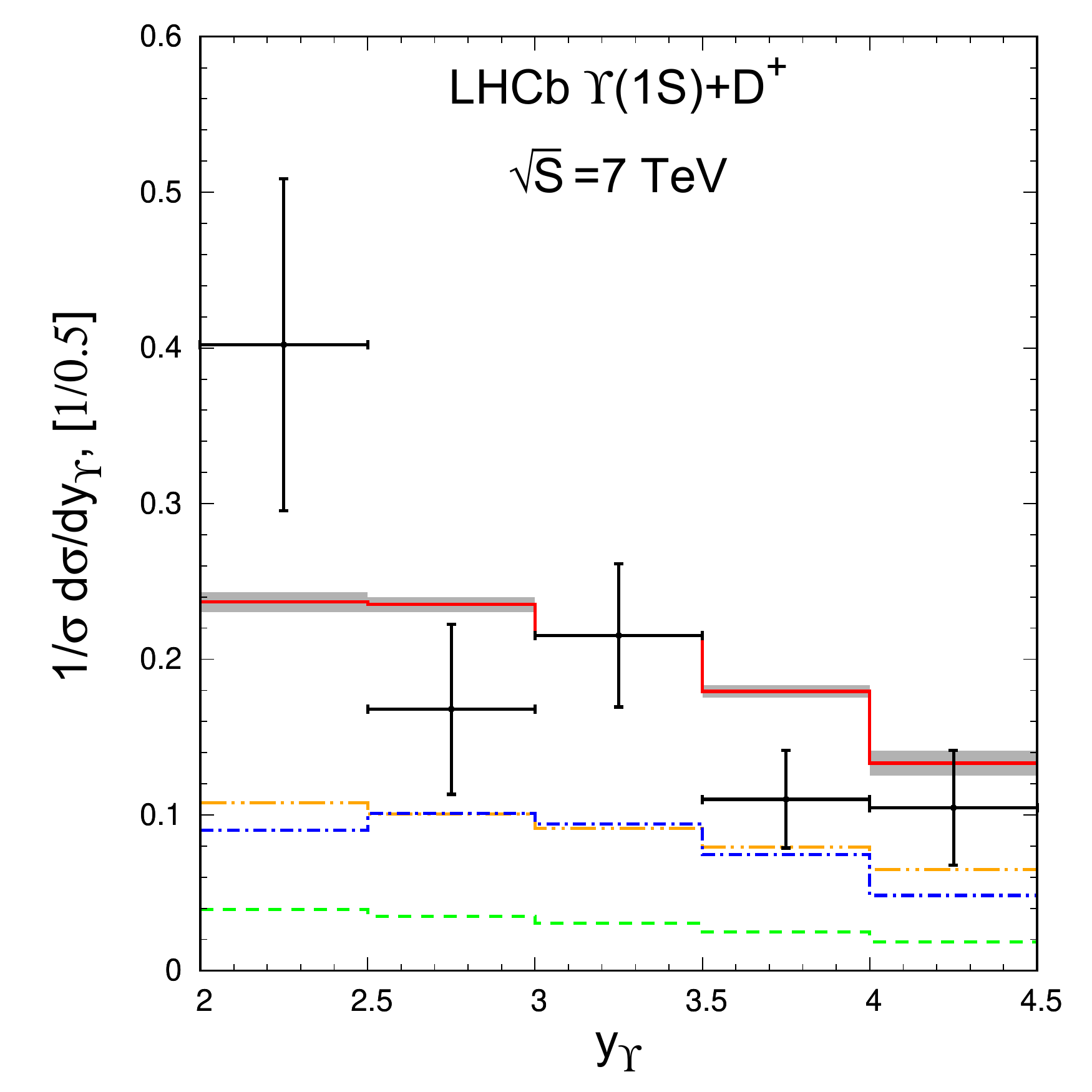}
\includegraphics[width=0.5\textwidth]{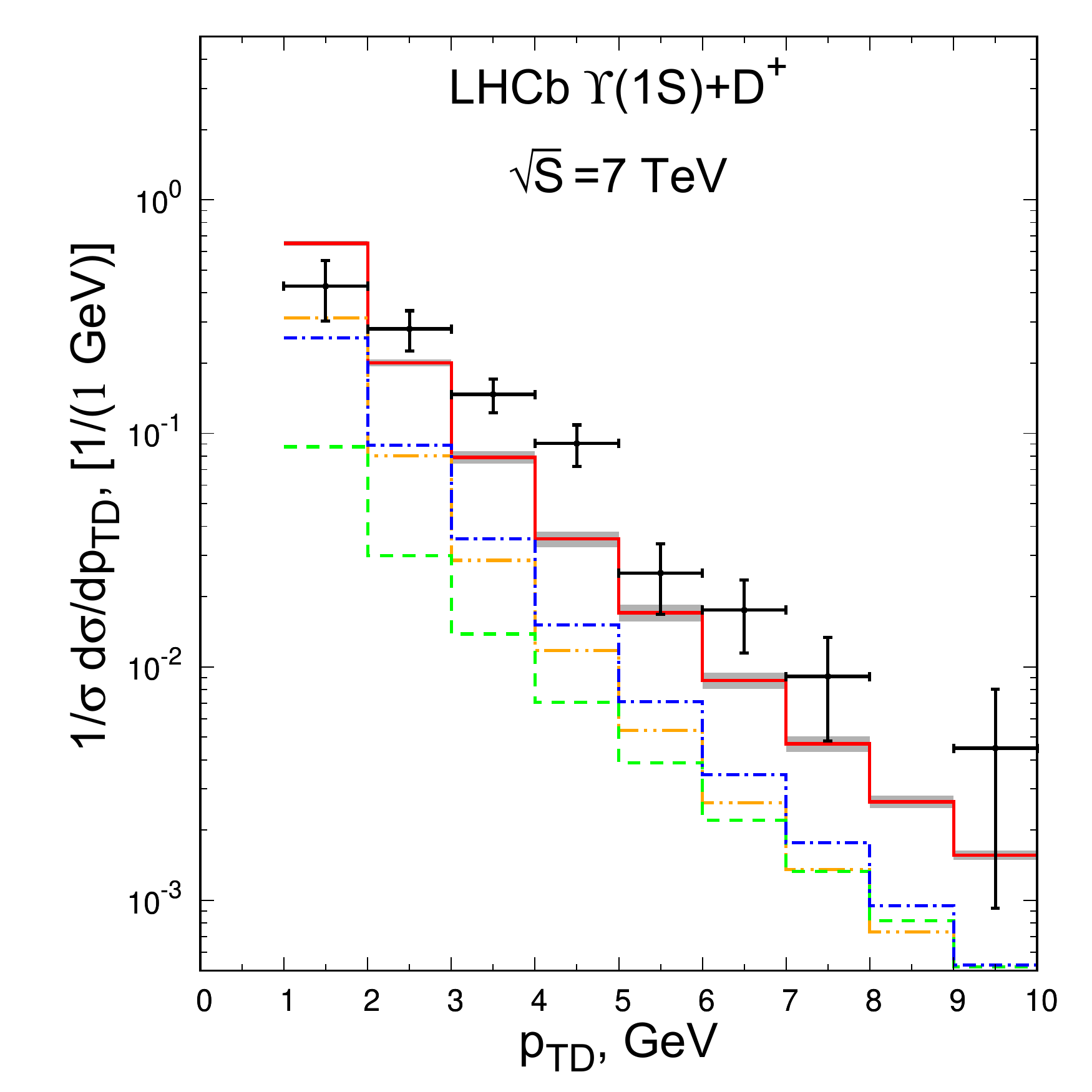}\includegraphics[width=0.5\textwidth ]{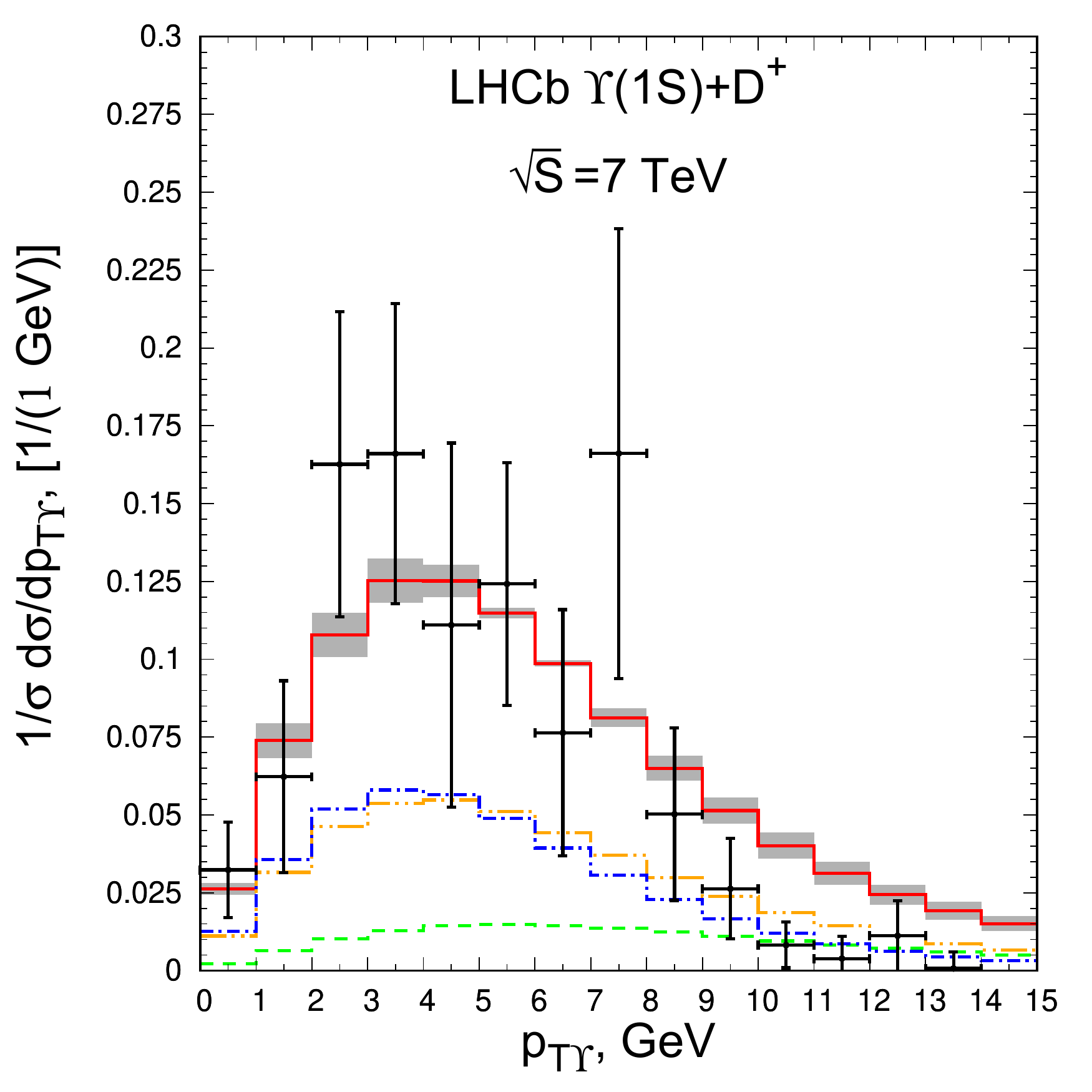}
 \caption{Transverse momentum and rapidity spectra of  $\Upsilon(1S)$ and $D^+$.
 The LHCb data~\cite{LHCb_ups_D} are taken at  $\sqrt{S}=7$ TeV, $2.0<y_{\Upsilon(D)}<4.5$, $0<p_{T\Upsilon}< 15$ GeV, and $1<p_{TD}< 20$ GeV.  Blue histograms are color-singlet
 contributions, green -- color-octet contributions, orange -- sum of feed-down contributions, red -- sum of all contributions.}
 \label{fig-1a}
\end{figure}

\begin{figure}[]
\includegraphics[width=0.5\textwidth  ]{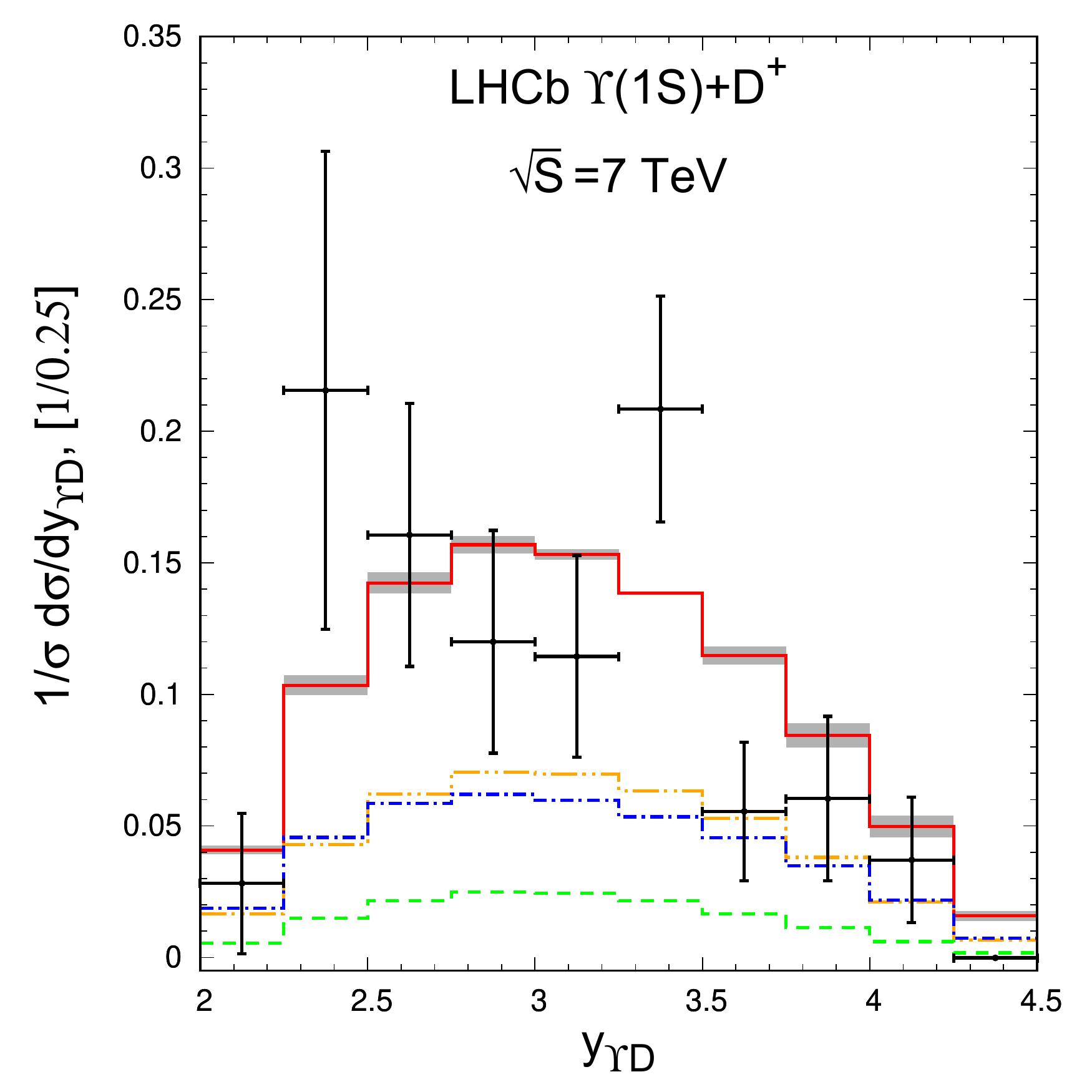}\includegraphics[width=0.5\textwidth  ]{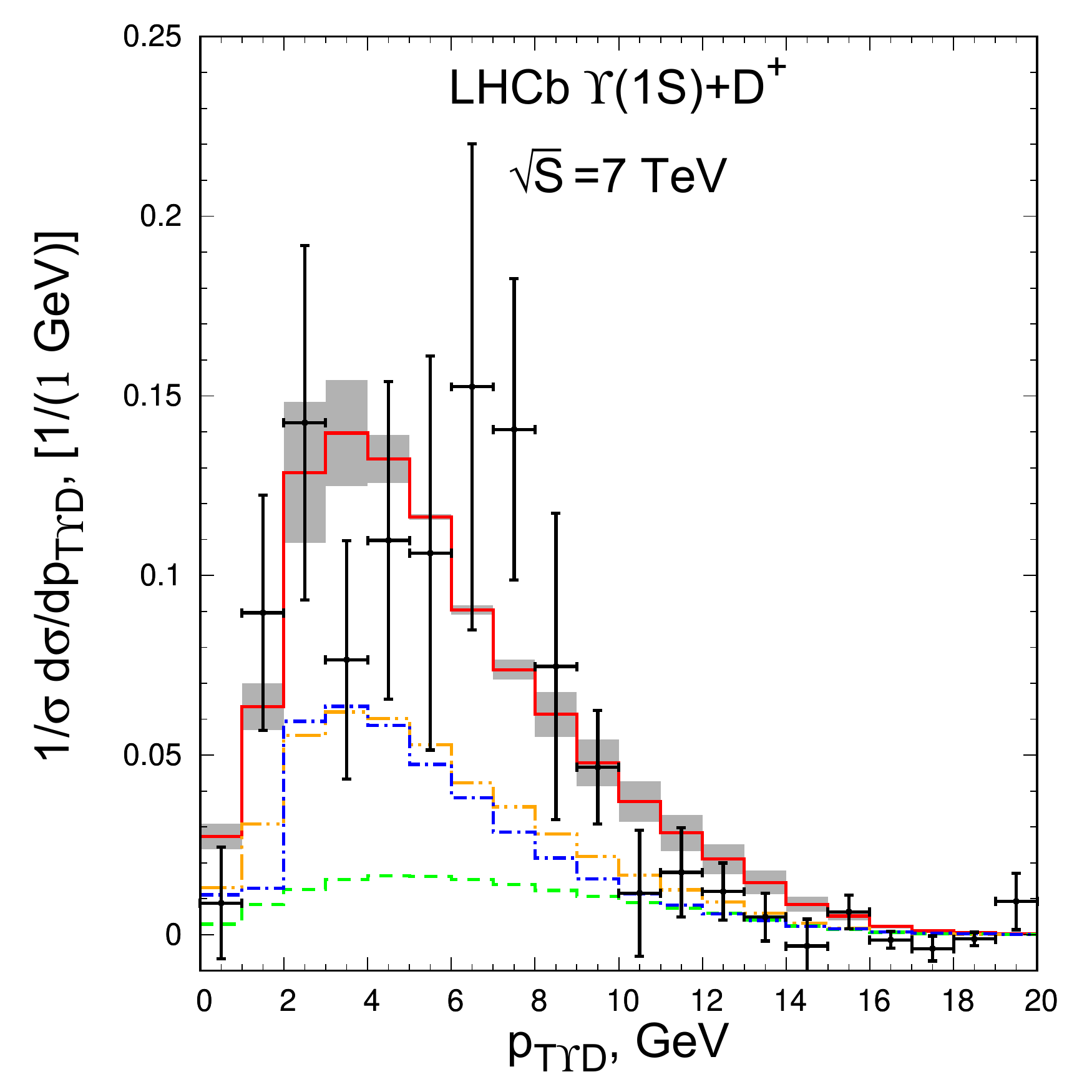}
\includegraphics[width=0.5\textwidth  ]{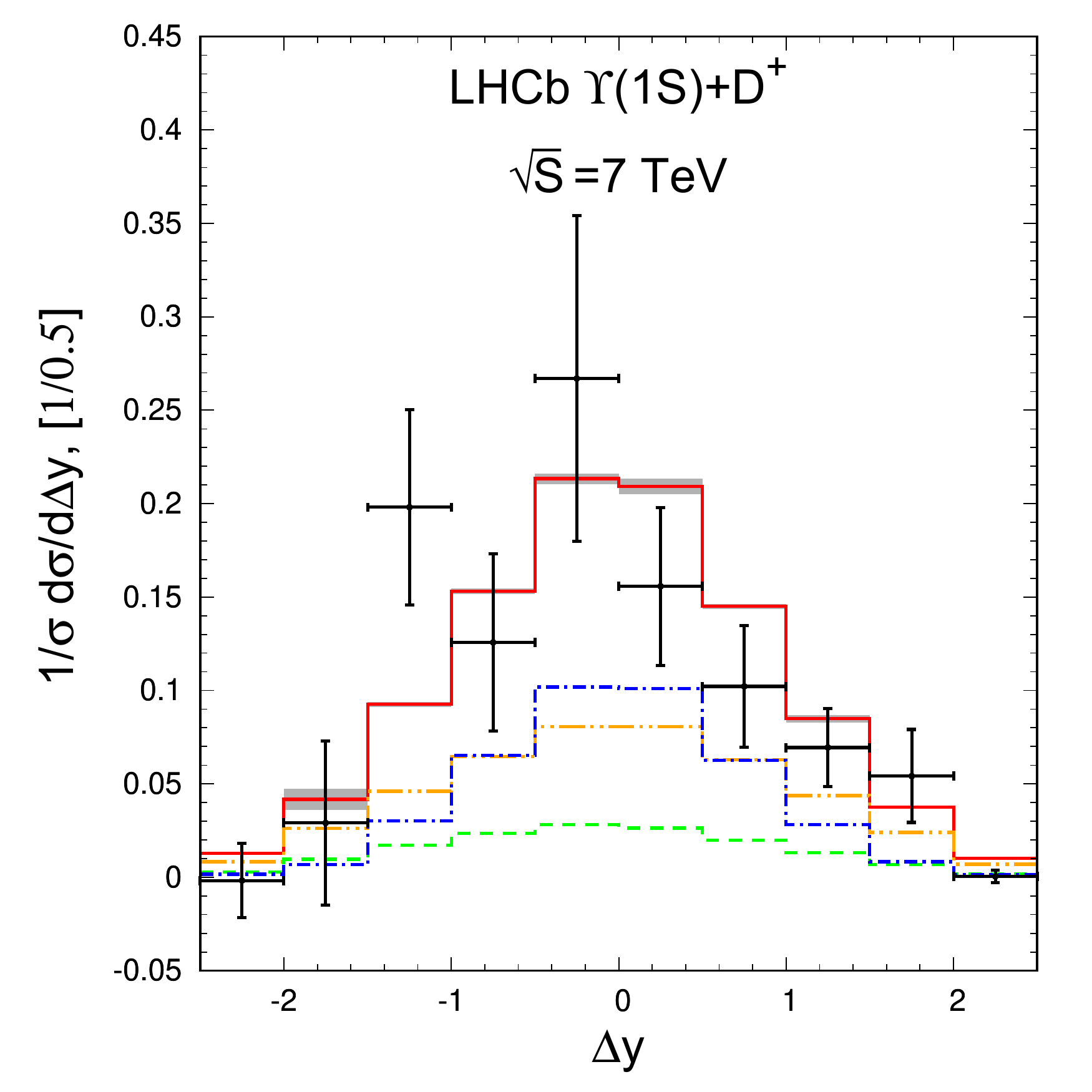}\includegraphics[width=0.5\textwidth  ]{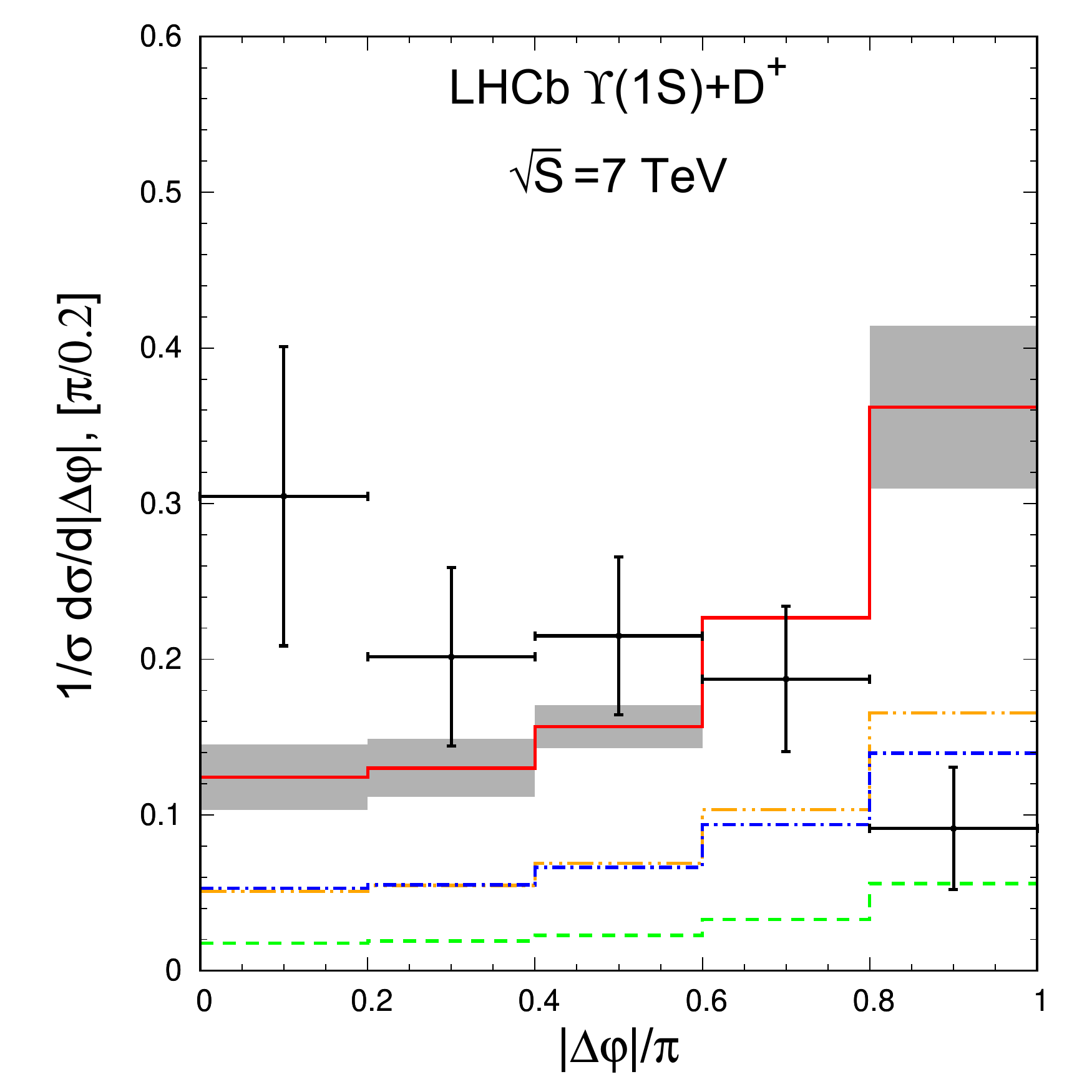}
 \caption{Top panel: transverse momentum and rapidity spectra of  $\Upsilon(1S)D^{+}$ pairs.
 Bottom panel: rapidity difference and azimuthal angle difference  spectra.  Histograms are the same as in the Fig. \ref{fig-1}. The LHCb data~\cite{LHCb_ups_D} are taken at
 $\sqrt{S}=7$ TeV, $2.0<y_{\Upsilon(D)}<4.5$, $0<p_{T\Upsilon}< 15$ GeV, and $1<p_{TD}< 20$ GeV.}
 \label{fig-2a}
\end{figure}

In the Figs. \ref{fig-1} and \ref{fig-1a}, we compare normalized
transverse momentum and rapidity spectra of $\Upsilon(1S)$ and
$D^{0,+}$ predicted by our model with LHCb data~\cite{LHCb_ups_D}.
The top panels of Figs. \ref{fig-2} and \ref{fig-2a} collect
rapidity and and transverse momentum spectra of $\Upsilon(1S)
D^{0,+}$ pairs. For all above-mentioned spectra, our hybrid LO PRA
plus NRQCD plus fragmentation model does reasonably well in
explaining the experimental data.

At the bottom panels of Figs. \ref{fig-2} and \ref{fig-2a} we plot
spectra for the rapidity difference $\Delta y=|y_{\Upsilon}-y_{D}|$
and for the azimuthal angle difference $\Delta
\varphi=|\varphi_{\Upsilon}-\varphi_{D}|$. The $\Delta y$ spectra
are reasonably well descried. The $\Delta\varphi$ spectrum obtained
in our model has typical shape for this kind of spectra. It has one
peak at the $\Delta\varphi\to \pi$ and plateau at the
$\Delta\varphi\leq \pi/2$. However, the experimental data from LHCb
Collaboration, though having large errors, demonstrate existence of
the second peak for $\Delta\varphi\to 0$. This feature of the data
can not be explained even by the DPS-model based on
Eq.~(\ref{Eq:pocket-formula}), which predicts the flat,
un-correlated $\Delta\varphi$-spectrum.

\begin{figure}[]
 \includegraphics[width=0.5\textwidth  ]{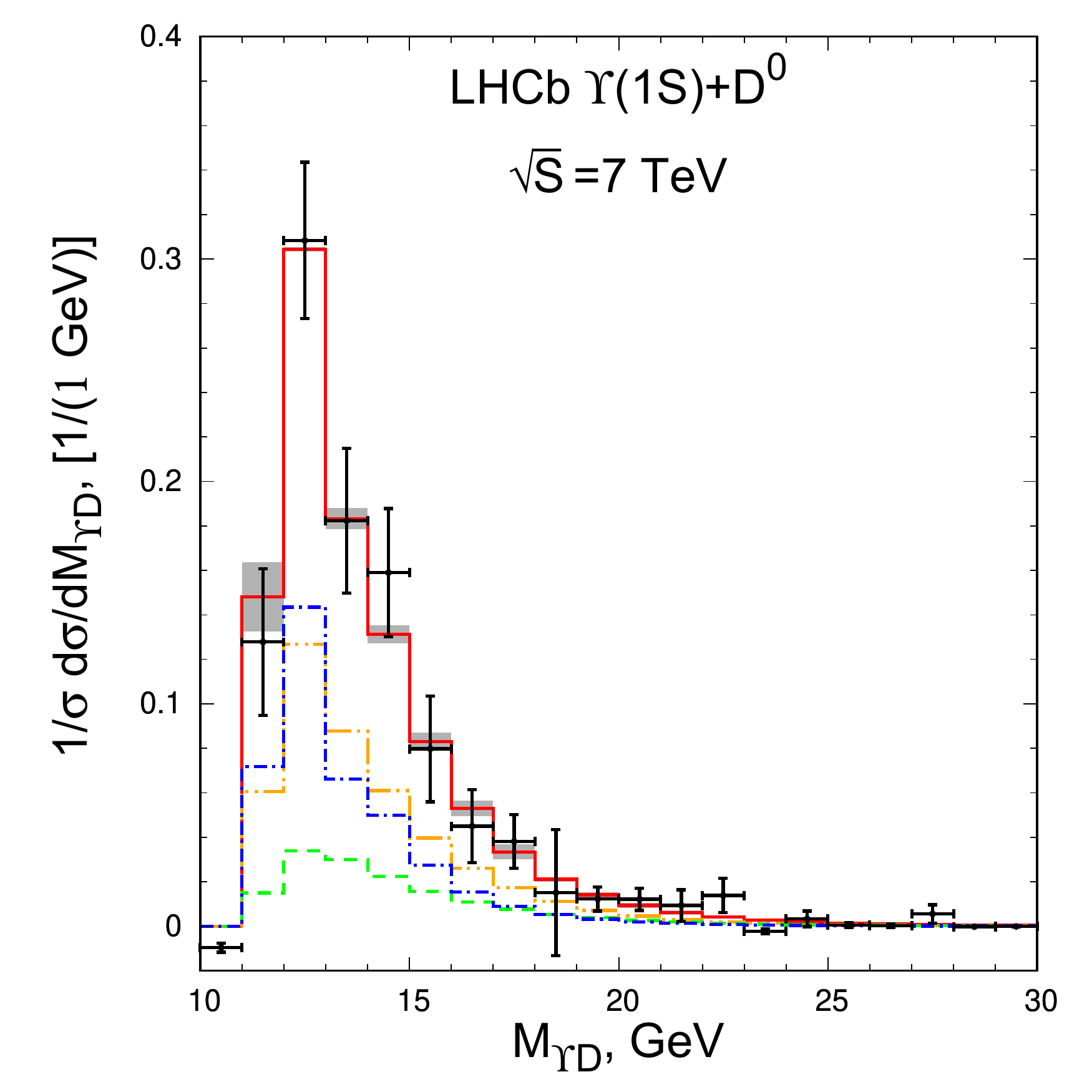}\includegraphics[width=0.5\textwidth  ]{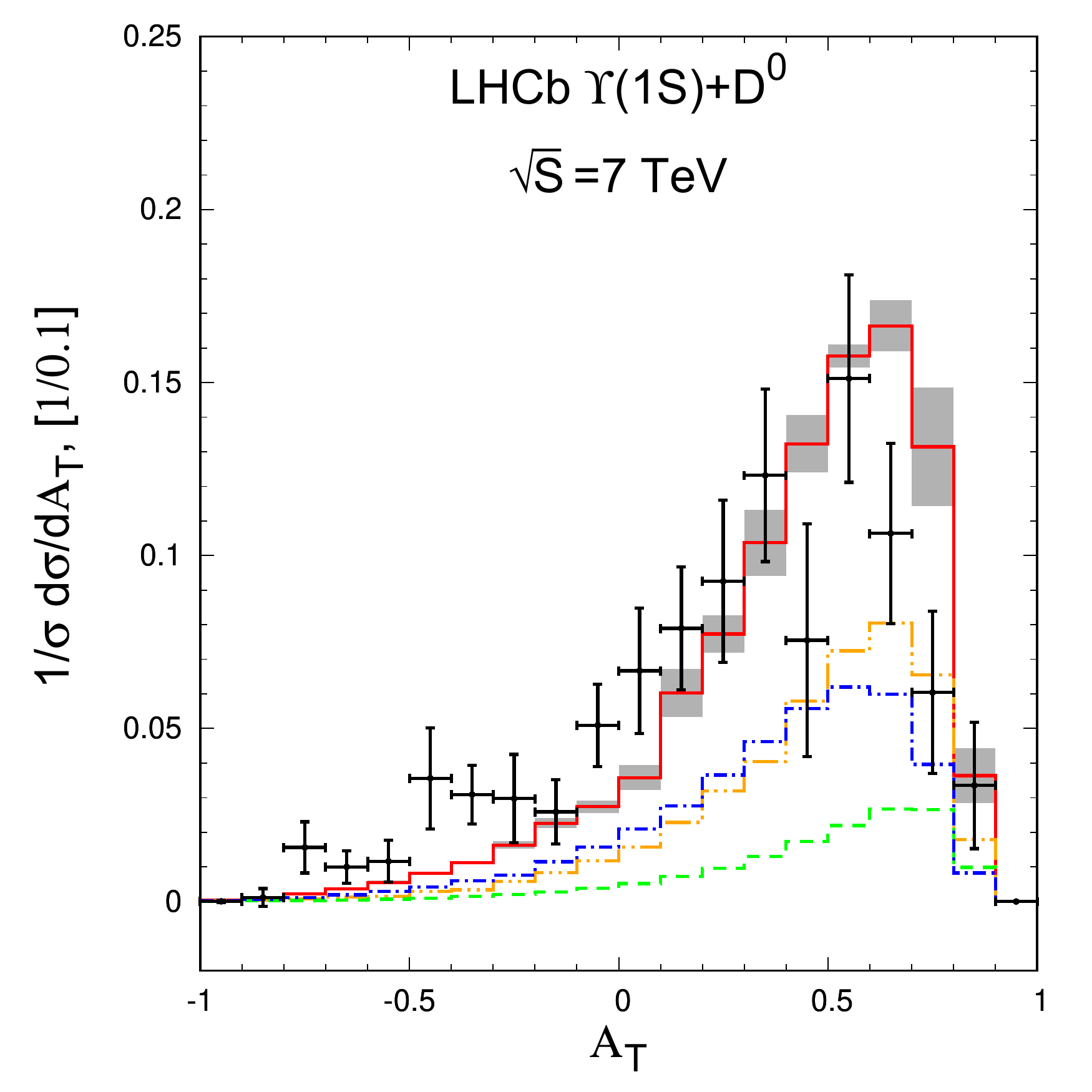}
 \caption{Left panel: invariant mass spectrum of  $\Upsilon(1S)D^0$ pairs.
 Right panel: transverse momentum asymmetry $\cal A_T$ spectrum . Histograms are the same as in the Fig.~\ref{fig-1}
 The LHCb data~\cite{LHCb_ups_D} are taken at $\sqrt{S}=7$ TeV, $2.0<y_{\Upsilon(D)}<4.5$, $0<p_{T\Upsilon}< 15$ GeV, and $1<p_{TD}< 20$ GeV.}
 \label{fig-3}
\end{figure}

\begin{figure}[]
 \includegraphics[width=0.5\textwidth  ]{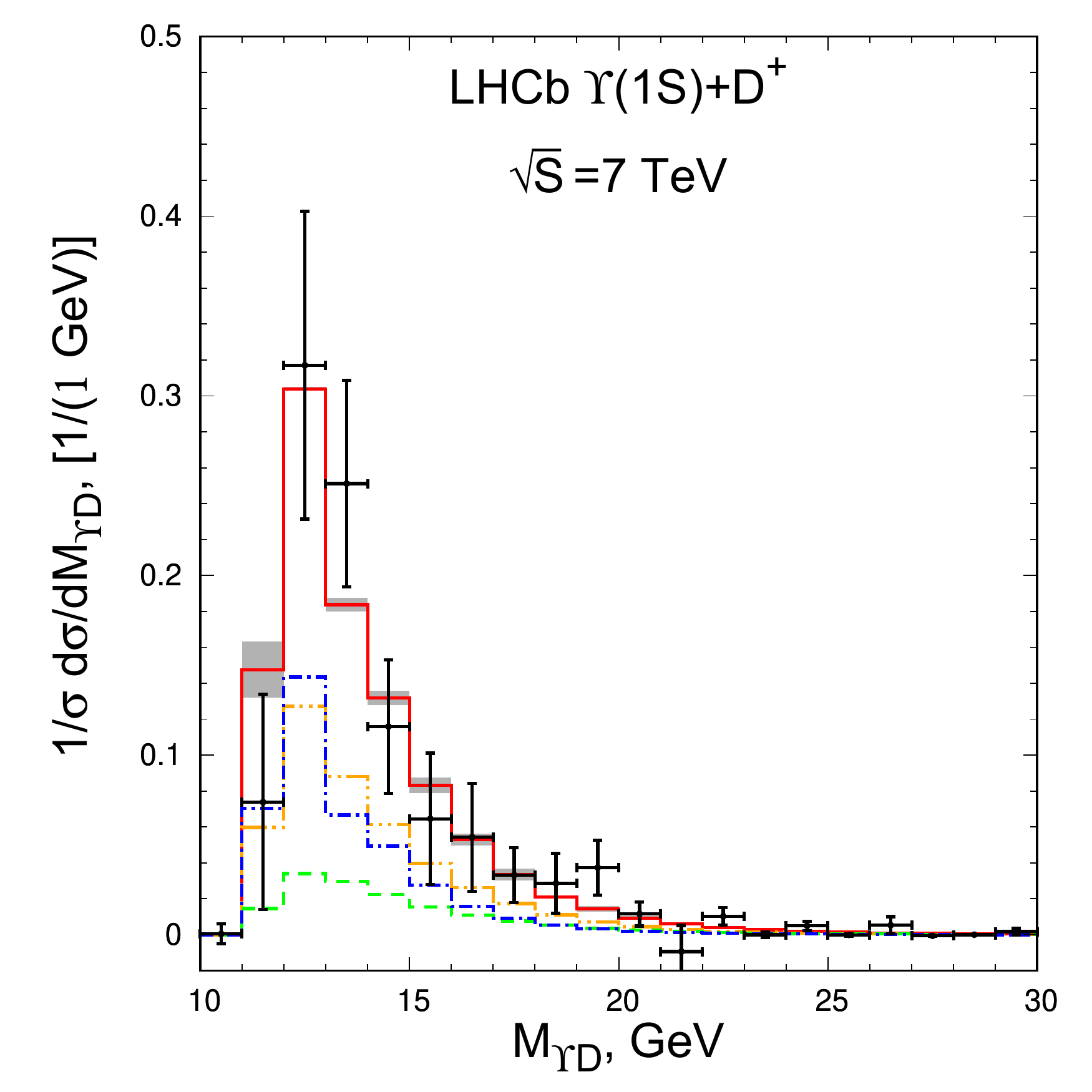}\includegraphics[width=0.5\textwidth  ]{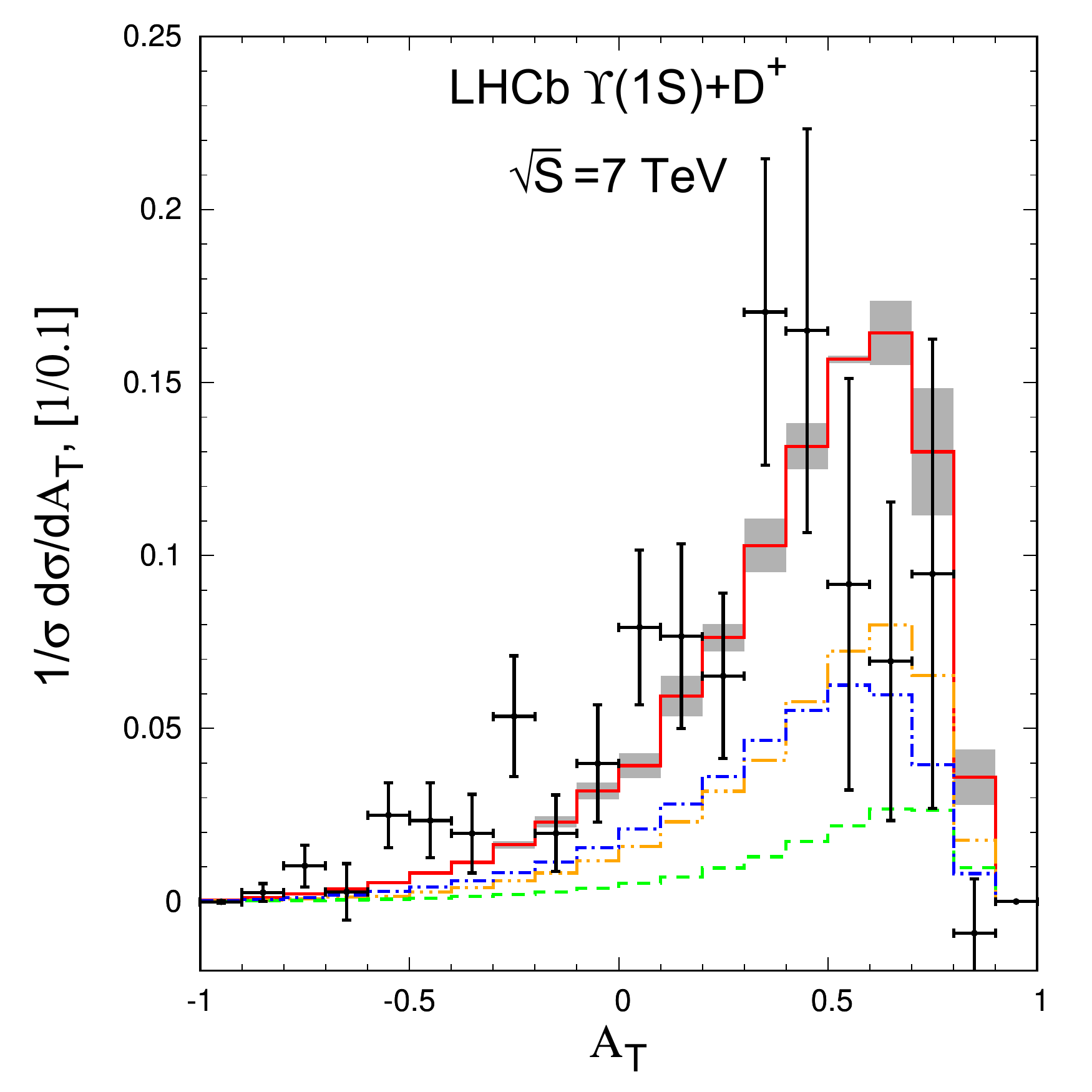}
 \caption{Left panel: invariant mass spectrum of  $\Upsilon(1S)D^+$ pairs.
 Right panel: transverse momentum asymmetry $\cal A_T$ spectrum . Histograms are the same as in the Fig.~\ref{fig-1}
 The LHCb data~\cite{LHCb_ups_D} are taken at $\sqrt{S}=7$ TeV, $2.0<y_{\Upsilon(D)}<4.5$, $0<p_{T\Upsilon}< 15$ GeV, and $1<p_{TD}< 20$ GeV.}
 \label{fig-3a}
\end{figure}

In the Figs. \ref{fig-3} and \ref{fig-3a}, we plot spectra for the
 invariant mass of
$\Upsilon(1S) D^{0,+}$ pair ($M$) and transverse momentum asymmetry
$A_T=(p_{T\Upsilon}-p_{TD})/(p_{T\Upsilon}+p_{TD})$. Predictions of our model with this correlation spectra also agree reasonably well with LHCb data.

Now we turn to the discussion of predictions of our SPS model for
total cross-sections $B_{\mu^+\mu^-}\times \sigma^{\Upsilon
D^{0,+}}$ (where $\Upsilon(1S)\to\mu^+\mu^-$ branching fraction
$B_{\mu^+\mu^-}=0.025$ is taken into account). At $\sqrt{S}=7$ TeV,
central values of our predictions for $\Upsilon(1S) D^{0}$ and
$\Upsilon(1S) D^{+}$ production are 91 pb and 36 pb respectively
(Tab.~\ref{TableII}) which, especially taking into account large
scale-uncertainty, reaches almost {more than} one half of the
experimental cross-section, which is respectively 155 and 82 pb.
Different contributions to the total cross section and it's
scale-uncertainty are presented in Tables \ref{TableII} and
\ref{TableIII}, correspondingly for the energies $\sqrt{S}=7$ TeV
and $\sqrt{S}=8$ TeV. Feed-down contribution from decays of higher
lying bottomonium states $(\Upsilon(3S), \Upsilon(2S), \chi_b(2P),
\chi_b(1P))$ is not small, it is about 50\% from total calculated
cross section. Contribution of color-singlet processes
(\ref{RRSwave}) and (\ref{RRPwave}) is always dominant, but
contribution of $[^3S_1^{(8)}]$ intermediate state is also very
important, reaching up to a half of the $[^3S_1^{(1)}]$ intermediate
state contribution.

In our model, the main source of $D^{0,+}$ mesons is the gluon
fragmentation. Let's compare above results with the model based on
$c-$quark fragmentation into $D$ mesons. We have calculated direct
production cross sections via color-singlet and color-octet
intermediate states in process (\ref{2-3Swave}) which on the quark
level corresponds to
\begin{equation}
R + R \to b \bar b[^3S_1^{(1,8)}]+c+\bar c.\label{eq-RRbbcc}
\end{equation}
Due to fragmentation model, we have to
take the $c-$quark as massless, however we take into account
threshold condition for invariant mass of $c\bar c$-pair:
$s_{c\bar{c}}=(p_c+p_{\bar c})^2>4m_c^2$ with $m_c=1.5$ GeV. Our
$2\to 3$ amplitudes passed numerous cross-checks, in particular, we
have numerically checked, that {\it final-state collinear} limit
($s_{c\bar{c}}\ll \min(p^2_{Tc},p^2_{T\bar{c}})$) of the squared
amplitudes of the processes (\ref{eq-RRbbcc}) is related with the
squared amplitudes of the processes (\ref{Ssinglet}, \ref{Soctet})
by the well-known collinear-factorization relation for tree-level
amplitudes:
\[
\left\langle\overline{|{\cal M}[RR\to b\bar{b}+c(p_c)+\bar{c}(p_{\bar{c}})]|^2}\right\rangle\simeq \frac{2g_s^2}{s_{c\bar{c}}} P_{qg}(z)\cdot \overline{|{\cal M}[RR\to b\bar{b}+g(p_c+p_{\bar{c}}))]|^2},
\]
where $z=p_c^0/(p_c^0+p_{\bar{c}}^0)$, $P_{qg}(z)=[z^2+(1-z)^2]/2$ and angular brackets in the l.h.s. stand for the averaging over azimuthal angle parametrizing the directions of momenta ${\bf p}_c$ and ${\bf p}_{\bar{c}}$ for constant ${\bf p}_c+{\bf p}_{\bar{c}}$ and $z$.

For the direct contributions from the subprocesses (\ref{eq-RRbbcc})
we obtain $B_{\mu^+\mu^-}\times \sigma^{D^0}_{\rm direct}[R+R\to
b\bar b[^3S_1^{(1+8)}]+c+\bar{c}]=20$ pb at $\sqrt{S}=7$ TeV and
$24$ pb at
 $\sqrt{S}=8$ TeV, so for ratio of direct cross sections we obtain:
\[
\frac{\sigma^{D^0}_{\rm direct}[R+R\to b\bar
b[^3S_1^{(1+8)}]+g]}{\sigma^{D^0}_{\rm direct}[R+R\to b\bar
b[^3S_1^{(1+8)}]+c+\bar c]}\simeq 2.6\div 2.5,
\]
at $\sqrt{S}=7$ and $8$ TeV. So our expectation of the dominant role of the gluon fragmentation in this process turns out to be correct.

\section{Conclusions}
\label{sec:conclusions} In the present paper we have demonstrated,
that conclusion about DPS-dominance in the total cross-section for
the process of associated hadroproduction of $\Upsilon$ and $D$
mesons, based on the LO CPM calculations of
Refs.~\cite{Likhoded2015,Likhoded2016} was premature and one can
relatively easily come up with the model which accounts {more than}
a half of the observed cross-section. Also, due to the use of
High-Energy factorization with unPDFs dependent on transverse
momenta of initial-state partons, the leading part of ISR
corrections has been taken into account and all differential
distributions are described reasonably well. Therefore one have to
be rather careful about the statements that DPS dominates total
cross-section of relatively low-scale, hadronization-sensitive
processes like one considered in the present paper. Most likely, the
solid evidence in favor of DPS can be obtained only from studies of
differential distributions in specific regions of phase space and at
high scales, where radiative corrections to SPS contribution can be
put under theoretical control.

\section*{Acknowledgments}
Authors thank the Ministry of Education and Science of the Russian
Federation for financial support in the framework of the Samara
University Competitiveness Improvement Program among the world's
leading research and educational centers for 2013-2020, the task No
3.5093.2017/8.9 and the Foundation for the Advancement of
Theoretical Physics and Mathematics BASIS, grant No. 18-1-1-30-1. We
thank Prof. B.A. Kniehl and Dr. Zhi-Guo He for stimulating
discussions.

\clearpage

\newpage

\begin{table}
 \caption{\label{TableI} The color-singlet and color-octet NMEs from Ref.~\cite{NSSUpsilon} used in the calculation.}
 \begin{tabular}{|c|c|}
 \hline
  NME & Fit in LO PRA. \\
  \hline
  $\left\langle {\cal O}^{\Upsilon(1S)}\left[^3S_1^{(1)}\right]\right\rangle\times$ GeV$^{-3}$ & $9.28$\\
  $\left\langle {\cal O}^{\Upsilon(1S)}\left[^3S_1^{(8)}\right]\right\rangle\times 10^2$ GeV$^{-3}$ & $2.31\pm 0.25$\\
  $\left\langle {\cal O}^{\Upsilon(1S)}\left[^1S_0^{(8)}\right]\right\rangle\times 10^2$ GeV$^{-3}$ & $0.0\pm 0.05$ \\
  $\left\langle {\cal O}^{\Upsilon(1S)}\left[^3P_0^{(8)}\right]\right\rangle\times 10^2$ GeV$^{-5}$ & $0.0\pm 0.38$ \\
 \hline
  $\left\langle {\cal O}^{\Upsilon(2S)}\left[^3S_1^{(1)}\right]\right\rangle\times$ GeV$^{-3}$ & $4.62$ \\
  $\left\langle {\cal O}^{\Upsilon(2S)}\left[^3S_1^{(8)}\right]\right\rangle\times 10^2$ GeV$^{-3}$ & $1.51\pm 0.17$ \\
  $\left\langle {\cal O}^{\Upsilon(2S)}\left[^1S_0^{(8)}\right]\right\rangle\times 10^2$ GeV$^{-3}$ & $0.0\pm 0.01$ \\
   $\left\langle {\cal O}^{\Upsilon(2S)}\left[^3P_0^{(8)}\right]\right\rangle\times 10^2$ GeV$^{-5}$ & $0.0\pm 0.03$ \\
  \hline
  $\left\langle {\cal O}^{\Upsilon(3S)}\left[^3S_1^{(1)}\right]\right\rangle\times$ GeV$^{-3}$ & $3.54$\\
  $\left\langle {\cal O}^{\Upsilon(3S)}\left[^3S_1^{(8)}\right]\right\rangle\times 10^2$ GeV$^{-3}$ & $1.24\pm 0.13$\\
  $\left\langle {\cal O}^{\Upsilon(3S)}\left[^1S_0^{(8)}\right]\right\rangle\times 10^2$ GeV$^{-3}$ & $0.0\pm 0.01$ \\
   $\left\langle {\cal O}^{\Upsilon(3S)}\left[^3P_0^{(8)}\right]\right\rangle\times 10^2$ GeV$^{-5}$ & $0.0\pm 0.02$\\
 \hline
  $\left\langle {\cal O}^{\chi(1P)}\left[^3P_0^{(1)}\right]\right\rangle\times$ GeV$^{-5}$ & $2.03$ \\
  $\left\langle {\cal O}^{\chi(1P)}\left[^3S_1^{(8)}\right]\right\rangle\times 10^2$ GeV$^{-3}$ & $0.0$ \\
 \hline
 $\left\langle {\cal O}^{\chi(2P)}\left[^3P_0^{(1)}\right]\right\rangle\times$ GeV$^{-5}$ & $2.36$ \\
  $\left\langle {\cal O}^{\chi(2P)}\left[^3S_1^{(8)}\right]\right\rangle\times 10^2$ GeV$^{-3}$ & $0.0$ \\
  \hline
 \end{tabular}
 \end{table}

\begin{table}
 \caption{\label{TableII} The total cross sections of $\Upsilon D^{+,0}$ production at the LHCb for $\sqrt{S}=7$ TeV.}
 \begin{ruledtabular}
 \begin{tabular}{|c|c|c|}
 & $B_{\mu^+\mu^-}\times
\sigma^{\Upsilon(1S)D^0}$, pb
& $B_{\mu^+\mu^-}\times \sigma^{\Upsilon(1S)D^+}$, pb \\
\hline
Direct contributions: & 51 & 20 \\
$R+R \to \Upsilon[^3S_1^{(1)}]+g$ & 37 & 15 \\
$R+R \to \Upsilon[^3S_1^{(8)}]+g$ &14 &  5 \\
\hline
Sum of feed-down contributions (\ref{RRSwave}), (\ref{RRPwave}) & 40& 16\\
\hline
Total cross section, LO PRA & $91^{+48}_{-41}$ & $36^{+19}_{-16}$\\
%

\hline Total cross section, experiment \cite{LHCb_ups_D} & $155 \pm
28$ & $82 \pm 24$
\end{tabular}
 \end{ruledtabular}
 \end{table}

\begin{table}
 \caption{\label{TableIII} The cross sections of $\Upsilon D^{+,0}$ production at the LHCb for $\sqrt{S}=8$ TeV.}
 \begin{ruledtabular}
 \begin{tabular}{|c|c|c|}
  & $B_{\mu^+\mu^-}\times
\sigma^{\Upsilon(1S)D^0}$, pb
& $B_{\mu^+\mu^-}\times \sigma^{\Upsilon(1S)D^+}$, pb \\
\hline
Direct contributions: & 61 & 24 \\
$R+R \to \Upsilon[^3S_1^{(1)}]+g$ & 44 & 18 \\
$R+R \to \Upsilon[^3S_1^{(8)}]+g$ &17 &  6 \\
\hline
Sum of feed-down contributions (\ref{RRSwave}), (\ref{RRPwave}) & 49& 19\\
\hline
Total cross section, LO PRA & $108^{+56}_{-48}$ & $42^{+22}_{-19}$\\
%

\hline Total cross section, experiment \cite{LHCb_ups_D} & $250 \pm
39$ & $80 \pm 21$
\end{tabular}
 \end{ruledtabular}
 \end{table}


%

\end{document}